\documentclass{aa}  

\usepackage{graphicx}
\usepackage{txfonts}
\usepackage{bm}
\usepackage{upgreek}
\usepackage{tikz}
\usepackage{environ}
\usepackage{algorithm}
\usepackage{algpseudocode}
\usepackage[inline]{enumitem}
\usepackage{booktabs}
\usepackage{makecell}
\usepackage{tablefootnote}
\usepackage{CJKutf8}

\newcommand{\fcfc}{\texttt{FCFC}}
\newcommand{\corrfunc}{\texttt{Corrfunc}}

\makeatletter
\newsavebox{\measure@tikzpicture}
\NewEnviron{scaletikzpicturetowidth}[1]{%
  \def\tikz@width{#1}%
  \begin{lrbox}{\measure@tikzpicture}%
  \BODY
  \end{lrbox}%
  \pgfmathparse{#1/\wd\measure@tikzpicture}%
  \BODY
}
\makeatother

\algnewcommand{\IIf}[1]{\State\algorithmicif\ #1\ \algorithmicthen}
\algnewcommand{\EndIIf}{\unskip\ \algorithmicend\ \algorithmicif}
\algnewcommand{\algorithmicor}{\textbf{ or }}
\algnewcommand{\OR}{\algorithmicor}

\makeatletter
\g@addto@macro{\UrlBreaks}{\UrlOrds}
\makeatother

\usetikzlibrary{shapes, decorations.pathreplacing, fadings, arrows.meta}

\definecolor{nodefgcolor}{RGB}{251,246,239}
\definecolor{nodebgcolor}{RGB}{255,239,159} 
\definecolor{nodebgcolor2}{RGB}{255,229,101} 
\definecolor{nodevisit}{RGB}{220,220,220}
\definecolor{memcolor}{RGB}{255,214,224}
\definecolor{pointcolor}{RGB}{255,50,100}
\definecolor{taborange}{RGB}{255,127,14}
\definecolor{tabred}{RGB}{214,39,40}

\tikzset{
    every picture/.append style={scale=0.4},
    pointstyle/.style = {mark=*, only marks, mark size=4},
    querypoint/.style = {pointcolor, very thick},
    boundary/.style = {line width=1pt},
    split0/.style = {line width=0.6pt},
    split1/.style = {line width=0.5pt},
    split2/.style = {line width=0.4pt},
    parentnode/.style = {dash pattern=on 2pt off 1pt, line width=0.4pt},
    leafnode/.style = {line width=0.6pt, text opacity=1},
    level 1/.style={sibling distance = 6cm, level distance = 2.5cm},
    level 2/.style={sibling distance = 3cm, level distance = 2.5cm},
    level 3/.style={sibling distance = 1.5cm, level distance = 2.5cm},
    treestyle/.style = {align=center, text centered, draw=black, text opacity=1},
    skippednode/.style = {treestyle, circle},
    visitednode/.style = {skippednode, fill=nodevisit},
    skippedleaf/.style = {treestyle, rectangle, minimum width=0.45cm, minimum height=0.45cm},
    visitedleaf/.style = {skippedleaf, fill=nodebgcolor},
    insideleaf/.style = {dash pattern=on 2.45pt off 2pt, line width=0.4pt},
    betweenleaf/.style = {line width=0.7pt},
    visitedmem/.style = {fill=memcolor, line width=0}
}

\begingroup\newif\ifmy
\IfFileExists{points.txt}{}{\mytrue}
\ifmy

\fi\endgroup

\begin{document}

\begin{CJK*}{UTF8}{gbsn}

   \title{Fast Correlation Function Calculator}

   \subtitle{A high-performance pair counting toolkit}
   
   \author{Cheng Zhao (赵成)
          \inst{\ref{inst:epfl}}
          }

   \institute{Institute of Physics, Laboratory of Astrophysics, \'Ecole Polytechnique F\'ed\'erale de Lausanne (EPFL), Observatoire de Sauverny, CH-1290 Versoix, Switzerland\\
              \email{\href{mailto:cheng.zhao@epfl.ch}{cheng.zhao@epfl.ch}}\label{inst:epfl}
             }

   \date{Received September 15, 1996; accepted March 16, 1997}

 
  \abstract
   {A novel high-performance exact pair counting toolkit called Fast Correlation Function Calculator (\fcfc{}) is presented, which is publicly available at \url{https://github.com/cheng-zhao/FCFC}.}
   {As the rapid growth of modern cosmological datasets, the evaluation of correlation functions with observational and simulation catalogues has become a challenge. High-efficiency pair counting codes are thus in great demand.}
   {We introduce different data structures and algorithms that can be used for pair counting problems, and perform comprehensive benchmarks to identify the most efficient ones for real-world cosmological applications.
    We then describe the three levels of parallelisms used by \fcfc{} -- including SIMD, OpenMP, and MPI -- and run extensive tests to investigate the scalabilities.
    Finally, we compare the efficiency of \fcfc{} against alternative pair counting codes.}
   {The data structures and histogram update algorithms implemented in \fcfc{} are shown to outperform alternative methods. \fcfc{} does not benefit much from SIMD as the bottleneck of our histogram update algorithm is mostly cache latency. Nevertheless, the efficiency of \fcfc{} scales well with the numbers of OpenMP threads and MPI processes, albeit the speedups may be degraded with over a few thousand threads in total.
   \fcfc{} is found to be faster than most (if not all) other public pair counting codes for modern cosmological pair counting applications.}
   {}


   \maketitle
%

\section{Introduction}

Correlation functions are a handy statistical tool in cosmology that characterises the excess probability of finding tracers with given separations compared to a random distribution.
Thus, they are a measure of the clustering pattern of a tracer distribution, which can then be used to infer statistical quantities of the underlying density field.
In the current standard cosmological paradigm, the distribution of matter results from tiny fluctuations in the primordial Universe, which evolve following gravitational instability and cosmic expansion.
For this reason, correlation functions are crucial for our understanding of inflation and cosmic structure formation models \citep[e.g.][]{Bernardeau2002}.
Pair correlation function -- also known as radial distribution function, which is essentially the isotropic 2-point correlation function (2PCF) -- is also a fundamental quantity in statistical mechanics that links microscopic details to macroscopic properties \citep[][]{Chandler1987}.

In fact, the measurement of 2PCFs of galaxies and quasars has been a key goal of massive spectroscopic surveys, such as Baryon Oscillation Spectroscopic Survey \citep[BOSS;][]{Dawson2013}, Extended Baryon Oscillation Spectroscopic Survey \citep[eBOSS;][]{Dawson2016}, and the ongoing Dark Energy Spectroscopic Instrument \citep[DESI;][]{DESI2016}.
With a data catalogue and the corresponding random sample, the 2PCF is generally measured using the Landy--Szalay (LS) estimator \citep[][]{Landy1993}:
\begin{equation}
\xi = ({\rm DD} - 2 {\rm DR} + {\rm RR}) / {\rm RR} ,
\end{equation}
where DD, DR, and RR denote the data--data, data--random, and random--random pair counts, respectively.
Nowadays, observational and simulated galaxy samples normally consist of millions or more galaxies.
Robust clustering measurements further require random samples with typically 10 times the objects.
As a result, the brute-force pair counting approach which evaluates $N^2$ pair separations -- where $N$ is the number of data points -- is impractical.
Actually, computing the 2PCFs from pair counts have become a practical challenge for modern cosmological analysis, not to mention higher-order statistics like 3-point correlation functions.

The evaluation of correlation functions is effectively a range searching problem, which reports objects within a query range.
Range searching is a fundamental topic in computational geometry. There are a variety of data structures and algorithms aim at solving range searching problems with different objects and query ranges \citep[see e.g.][]{deBerg2008}.
In the context of cosmology, efficient correlation function calculators have also been studied extensively in literature, from the pioneering work by \citet[][]{Moore2001} to the recent remarkable development of \citet[][]{CORRFUNC}.
Meanwhile, there are significant efforts on making full use of high-performance computing (HPC) resources, such as a large number of multi-core CPUs and GPUs \citep[e.g.][]{Dolence2008,CUTE,Chhugani2012,Ponce2012}.
Different approximate methods are explored widely as well \citep[e.g.][]{Zhang2005,Slepian2015,Philcox2022}.

Despite the large number of publicly available pair counting codes on the market \citep[e.g.][and references therein]{CUTE,TREECORR,correlcalc,GUNDAM,CORRFUNC}, we introduce Fast Correlation Function Calculator\footnote{\url{https://github.com/cheng-zhao/FCFC}} (\fcfc{}), a novel high-efficiency, scalable, portable, flexible, and user-friendly toolkit for exact pair counting.
We focus on \fcfc{} version 1.0.1 in this article, which supports 2PCFs for 3D data, with various commonly used binning schemes.
It is possibly the fastest publicly available 2PCF calculator for modern cosmological datasets so far.

This paper is organised as follows.
We begin with a comparison of different data structures for pair counting problems in Sect.~\ref{sec:data_struct}.
Then, we introduce the pair counting and histogram update algorithms used by \fcfc{} in Sect.~\ref{sec:alg}.
In Sect.~\ref{sec:parallel}, we describe the performance of \fcfc{} with different levels of parallelisms.
A direct comparison between \fcfc{} and \corrfunc{}, another efficient cosmological pair counting code, is presented in Sect.~\ref{sec:comp_corrfunc}.
Finally, we conclude in Sect.~\ref{sec:conclusion}.


\section{Data structures}
\label{sec:data_struct}

Data structures are a technique to organise and store an input dataset in memory that allows efficient data access. Typically, it is not only unnecessary, but also inefficient to process the data all at once in a program. A well-designed data structure may prevent the retrieval of irrelevant data during data queries as much as possible, which is known as data pruning. Thus, data structures are usually crucial for efficient algorithms.
To this end, several types of data structures for pair counting applications have been proposed in literature, including regular grids \citep[][]{CUTE, CORRFUNC}, linked list \citep[][]{GUNDAM}, $k$-d tree \citep[][]{Moore2001}, and ball tree \citep[][]{correlcalc}.

\begin{figure}
\centering
\begin{tikzpicture}
\draw[boundary, rounded corners, fill=nodebgcolor] (2,6.5) rectangle (16,10.5);
\node at (9,9.8) {\sf Abstract data type};
\draw[leafnode, fill=nodefgcolor] (2.5,7) rectangle (6.5,9);
\node[text width=1.7cm, align=center] at (4.5,8) {\small data summary};
\draw[leafnode, fill=nodefgcolor] (7,7) rectangle (11,9);
\node[text width=1.5cm, align=center] at (9,8) {\small data pointer};
\draw[parentnode, fill=nodefgcolor] (11.5,7) rectangle (15.5,9);
\node[text width=1.7cm, align=center] at (13.5,8) {\small link to other nodes};
\draw[betweenleaf] (0.5,4.5) -- (0.5,5.5) -- (17.5,5.5) -- (17.5,4.5);
\draw[betweenleaf] (5.5,4.5) -- (5.5,5.5);
\draw[betweenleaf] (12.5,4.5) -- (12.5,5.5);
\draw[boundary] (9,5.5) -- (9,6.5);
\foreach \x in {0, 1} {
    \draw[leafnode, rounded corners, fill=nodebgcolor] (5*\x-1,3) rectangle (5*\x+2,4.5);
    \draw[leafnode, rounded corners, fill=nodebgcolor] (16-5*\x,3) rectangle (19-5*\x,4.5);
}
\draw[split0, dash pattern=on 2pt off 1.5pt] (2,3.75) -- (4,3.75);
\draw[split0, dash pattern=on 2pt off 1.5pt] (7,3.75) -- (7.8,3.75);
\draw[split0, dotted] (8.4,3.75) -- (9.6,3.75);
\node at (9,4.5) {\sf Nodes};
\draw[split0, dash pattern=on 2pt off 1.5pt] (10.2,3.75) -- (11,3.75);
\draw[split0, dash pattern=on 2pt off 1.5pt] (14,3.75) -- (16,3.75);
\foreach \x in {0, 1} {
    \draw[split0, decorate, decoration={brace,raise=1.5pt,amplitude=4pt}] (3*\x,1) -- (3*\x+3,1);
    \draw[split0, decorate, decoration={brace,raise=1.5pt,amplitude=4pt}] (15-3*\x,1) -- (18-3*\x,1);
}
\draw[-stealth, split0] (0.5,3) -- (0.5,2.4) -- (1.5,2.4) -- (1.5,1.6);
\draw[-stealth, split0] (5.5,3) -- (5.5,2.4) -- (4.5,2.4) -- (4.5,1.6);
\draw[-stealth, split0] (12.5,3) -- (12.5,2.4) -- (13.5,2.4) -- (13.5,1.6);
\draw[-stealth, split0] (17.5,3) -- (17.5,2.4) -- (16.5,2.4) -- (16.5,1.6);
\fill[visitedmem] (0,0) rectangle (6,1);
\fill[visitedmem] (12,0) rectangle (18,1);
\shade[left color=memcolor, right color=white] (6,0) rectangle (7.5,1);
\shade[left color=white, right color=memcolor] (10.5,0) rectangle (12,1);
\draw[boundary] (6.5,0) -- (0,0) -- (0,1) -- (6.5,1);
\draw[boundary] (11.5,0) -- (18,0) -- (18,1) -- (11.5,1);
\draw[boundary, dashed] (6.5,0) -- (11.5,0);
\draw[boundary, dashed] (6.5,1) -- (11.5,1);
\foreach \x in {0, 1} {
  \draw[betweenleaf] (3*\x+3,0) -- (3*\x+3,1);
  \draw[insideleaf] (3*\x+1,0) -- (3*\x+1,1);
  \draw[insideleaf] (3*\x+2,0) -- (3*\x+2,1);
  \draw[betweenleaf] (15-3*\x,0) -- (15-3*\x,1);
  \draw[insideleaf] (17-3*\x,0) -- (17-3*\x,1);
  \draw[insideleaf] (16-3*\x,0) -- (16-3*\x,1);
}
\node at (9,-0.8) {\sf Data in memory};
\end{tikzpicture}
\caption{The architecture of data structures implemented in this work.}
\label{fig:data_struct}
\end{figure}
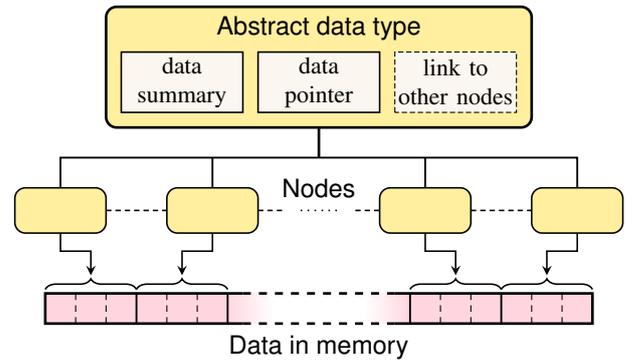

In general, a data structure sorts and partitions the dataset, and store the data segments on different nodes, either by copying the data directly, or saving only the addresses in memory.
Each node is typically defined as an abstract data type, which contains summaries of the associate data, though sometimes implicitly, for quickly judging whether the data should be retrieved.
Connections between different nodes may also be built, to optimise the node traversal process.
This architecture is illustrated in Fig.~\ref{fig:data_struct}.
Given the large datasets for cosmological applications, we save only data pointers on the nodes, to make the latter more compact, and reduce the chance of cache misses during node traversal.
The raw data, which are accessed less often, are stored separately and continuously in the memory.
In particular, during the construction of the data structures, the data is sorted in a way that points belonging to adjacent nodes are aligned continuously.

For pair counting applications, it is crucial to be able to compute the separation ranges between nodes efficiently, to omit nodes that are too far away or too close to each other, without visiting individual data points.
For this purpose, we describe a few data structures in this section -- including regular grids, $k$-d tree, and a new variant of the ball tree -- and compare their performances in terms of pair counting.
Note that throughout this section, the costs of structure constructions are not counted for our benchmarks, as they generally take $<1$ per cent of the time used by the pair counting processes.
Moreover, the computational costs are all measured with a single \textsf{Haswell} CPU core (see Appendix~\ref{sec:benchmark_setting} for details).

\subsection{Regular grids}
\label{sec:struct_grid}

A simple way to partition a dataset is to divide the domain into regular axis-aligned grids, with unique identifiers for spatial indexing.
Normally the positions and extents of the grid cells can be expressed by polynomials of the identifiers, or indices. 
Therefore, distance ranges between different grid cells can be inferred from the differences of cell indices, which can be computed prior to the cell traversal process. This makes regular grids a potentially very efficient data structure for pair counting.

With the architecture shown in Fig.~\ref{fig:data_struct}, only three passes through the dataset are required to construct regular grids for a catalogue:
\begin{enumerate*}[label=(\arabic*), itemjoin={{, }}, itemjoin*={{, and }}, labelwidth=0pt]
\item
find the minimum axis-aligned bounding box (AABB) of the catalogue to define grids
\item
count the number of data points in each grid cell
\item
group data points based on the associate cell indices.
\end{enumerate*}
Therefore, the construction of regular grids can be very efficient, with a time complexity of $\mathcal{O}(N)$, where $N$ denotes the total number of data points.
In contrast, the storage consumed by regular grids is very sensitive to the number of grid cells, and scales as $\mathcal{O}(\prod_i N_{{\rm g}, i})$, where $N_{{\rm g}, i}$ indicates the number of cells along the $i$-th dimension.

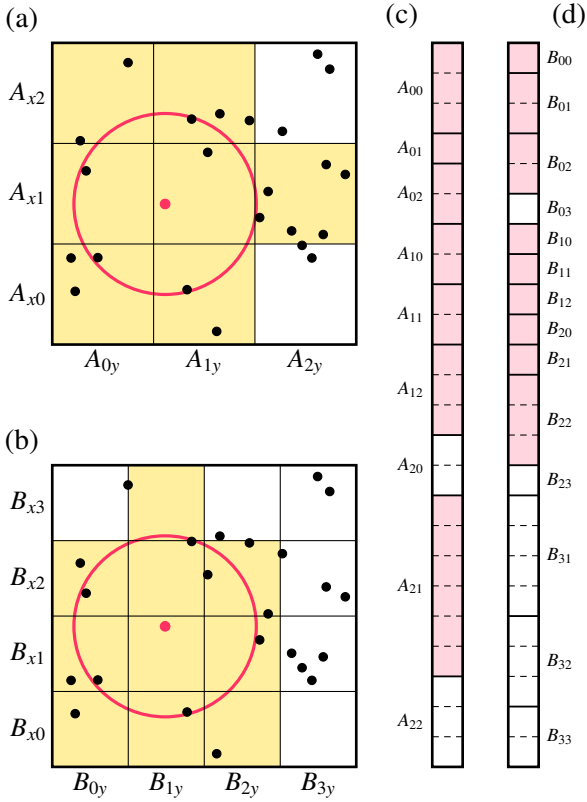
\begin{figure}
\centering
\begin{tikzpicture}
\begin{scope}
\fill[nodebgcolor] (0,0) rectangle (20/3,10);
\fill[nodebgcolor] (20/3,10/3) rectangle (10,20/3);
\draw[querypoint] (3.7123921,4.6591658) circle [radius=3];
\foreach \x in {0, 1, 2} {
    \draw[split2] (10/3*\x,0) -- (10/3*\x,10);
    \draw[split2] (0,10/3*\x) -- (10,10/3*\x);
    \node at (10/3*\x+10/6,-0.6) {$A_{\x y}$};
    \node at (-0.8,10/3*\x+10/6) {$A_{x \x}$};
}
\draw plot[pointstyle] file {points.txt};
\filldraw[querypoint] (3.7123921,4.6591658) circle [radius=3.5pt];
\draw[boundary] (0,0) rectangle (10,10);
\node[font=\large, align=right, below] at (-1,11.5) {(a)};
\end{scope}
\begin{scope}[yshift=-14cm]
\fill[nodebgcolor] (0,0) rectangle (7.5,7.5);
\fill[nodebgcolor] (2.5,7.5) rectangle (5,10);
\draw[querypoint] (3.7123921,4.6591658) circle [radius=3];
\foreach \x in {0,...,3} {
    \draw[split2] (2.5*\x,0) -- (2.5*\x,10);
    \draw[split2] (0,2.5*\x) -- (10,2.5*\x);
    \node at (2.5*\x+1.25,-0.6) {$B_{\x y}$};
    \node at (-0.8,2.5*\x+1.25) {$B_{x \x}$};
}
\draw plot[pointstyle] file {points.txt};
\filldraw[querypoint] (3.7123921,4.6591658) circle [radius=3.5pt];
\draw[boundary] (0,0) rectangle (10,10);
\node[font=\large, align=right, below] at (-1,11.5) {(b)};
\end{scope}
\begin{scope}[xshift=12.5cm, yshift=-14cm]
\fill[visitedmem] (0,11) rectangle (1,24);
\fill[visitedmem] (0,3) rectangle (1,9);
\draw[boundary] (0,0) rectangle (1,24);
\draw[insideleaf] (0,23) -- (1,23);
\draw[insideleaf] (0,22) -- (1,22);
\draw[betweenleaf] (0,21) -- (1,21);
\node at (-0.7,22.5) {$\scriptstyle A_{00}$};
\draw[betweenleaf] (0,20) -- (1,20);
\node at (-0.7,20.5) {$\scriptstyle A_{01}$};
\draw[insideleaf] (0,19) -- (1,19);
\draw[betweenleaf] (0,18) -- (1,18);
\node at (-0.7,19) {$\scriptstyle A_{02}$};
\draw[insideleaf] (0,17) -- (1,17);
\draw[betweenleaf] (0,16) -- (1,16);
\node at (-0.7,17) {$\scriptstyle A_{10}$};
\draw[insideleaf] (0,15) -- (1,15);
\draw[betweenleaf] (0,14) -- (1,14);
\node at (-0.7,15) {$\scriptstyle A_{11}$};
\draw[insideleaf] (0,13) -- (1,13);
\draw[insideleaf] (0,12) -- (1,12);
\draw[betweenleaf] (0,11) -- (1,11);
\node at (-0.7,12.5) {$\scriptstyle A_{12}$};
\draw[insideleaf] (0,10) -- (1,10);
\draw[betweenleaf] (0,9) -- (1,9);
\node at (-0.7,10) {$\scriptstyle A_{20}$};
\foreach \x in {0,...,4}
    \draw[insideleaf] (0,8-\x) -- (1,8-\x);
\draw[betweenleaf] (0,3) -- (1,3);
\node at (-0.7,6) {$\scriptstyle A_{21}$};
\draw[insideleaf] (0,2) -- (1,2);
\draw[insideleaf] (0,1) -- (1,1);
\node at (-0.7,1.5) {$\scriptstyle A_{22}$};
\node[font=\large, align=right] at (-1,24.9) {(c)};
\end{scope}
\begin{scope}[xshift=15cm, yshift=-14cm]
\fill[visitedmem] (0,19) rectangle (1,24);
\fill[visitedmem] (0,10) rectangle (1,18);
\draw[boundary] (0,0) rectangle (1,24);
\draw[betweenleaf] (0,23) -- (1,23);
\node at (1.7,23.5) {$\scriptstyle B_{00}$};
\draw[insideleaf] (0,22) -- (1,22);
\draw[betweenleaf] (0,21) -- (1,21);
\node at (1.7,22) {$\scriptstyle B_{01}$};
\draw[insideleaf] (0,20) -- (1,20);
\draw[betweenleaf] (0,19) -- (1,19);
\node at (1.7,20) {$\scriptstyle B_{02}$};
\draw[betweenleaf] (0,18) -- (1,18);
\node at (1.7,18.5) {$\scriptstyle B_{03}$};
\draw[betweenleaf] (0,17) -- (1,17);
\node at (1.7,17.5) {$\scriptstyle B_{10}$};
\draw[betweenleaf] (0,16) -- (1,16);
\node at (1.7,16.5) {$\scriptstyle B_{11}$};
\draw[betweenleaf] (0,15) -- (1,15);
\node at (1.7,15.5) {$\scriptstyle B_{12}$};
\draw[betweenleaf] (0,14) -- (1,14);
\node at (1.7,14.5) {$\scriptstyle B_{20}$};
\draw[betweenleaf] (0,13) -- (1,13);
\node at (1.7,13.5) {$\scriptstyle B_{21}$};
\draw[insideleaf] (0,12) -- (1,12);
\draw[insideleaf] (0,11) -- (1,11);
\draw[betweenleaf] (0,10) -- (1,10);
\node at (1.7,11.5) {$\scriptstyle B_{22}$};
\draw[betweenleaf] (0,9) -- (1,9);
\node at (1.7,9.5) {$\scriptstyle B_{23}$};
\draw[insideleaf] (0,8) -- (1,8);
\draw[insideleaf] (0,7) -- (1,7);
\draw[insideleaf] (0,6) -- (1,6);
\draw[betweenleaf] (0,5) -- (1,5);
\node at (1.7,7) {$\scriptstyle B_{31}$};
\draw[insideleaf] (0,4) -- (1,4);
\draw[insideleaf] (0,3) -- (1,3);
\draw[betweenleaf] (0,2) -- (1,2);
\node at (1.7,3.5) {$\scriptstyle B_{32}$};
\draw[insideleaf] (0,1) -- (1,1);
\node at (1.7,1) {$\scriptstyle B_{33}$};
\node[font=\large, align=right] at (2,24.9) {(d)};
\end{scope}
\end{tikzpicture}
\caption{Illustration of isotropic range searching using regular grids with different cell sizes. The points in (a) and (b) indicate a randomly generated dataset in 2D, with the current reference point marked in red. Yellow areas denote cells that are visited, given the query range indicated by red circles. Panels (c) and (d) show the arrangements of data points in memory, for the column-major grid configurations in (a) and (b) respectively. Pink regions indicate data points that are visited during the range searching process.}
\label{fig:grid}
\end{figure}

The efficiency of data pruning for regular grids depends largely on the cell sizes as well. An example is shown in Fig.~\ref{fig:grid}, where the data partitions with two different cell sizes are illustrated.
Given the same reference point and maximum distance for an isotropic range searching, the numbers of visited cells and data points are both significantly different when varying the number of cells per box side.
Here, data points belonging to different grid cells are arranged in column-major order, and gaps between adjacent memory visits are observed.
Sorting the cells using a space filling curve, such as the Hilbert curve, may improve the memory locality and reduce the chance of cache misses \citep[see e.g.][for an application]{Springel2005}. 
Nevertheless, the improvement is expected to be marginal, as the memory jumps can never be entirely eliminated, and it is more difficult, though possible, to pre-compute the map from indices of grid cells to the distance ranges between cells.

The algorithm for pair counting with regular grids is as simple as traversing all grid cells that contain data points, and visit successively cells that are separated within the distance range of interest, given the pre-computed offsets of indices.
Consequently, the complexity of the algorithm depends not only on the number of grid cells that intersect with the query range, but also on the average number of data points in each cell. Apparently, when increasing the side lengths of regular grid cells, the number of cells to be visited are reduced, but there may be more unnecessary distance evaluations for pairs, as illustrated by Fig.~\ref{fig:grid}. Thus, the choice of cell sizes is crucial for the efficiency of a grid-based pair counting algorithm \citep[see also][for relevant discussions]{CORRFUNC}.

\begin{figure}
\centering
\includegraphics{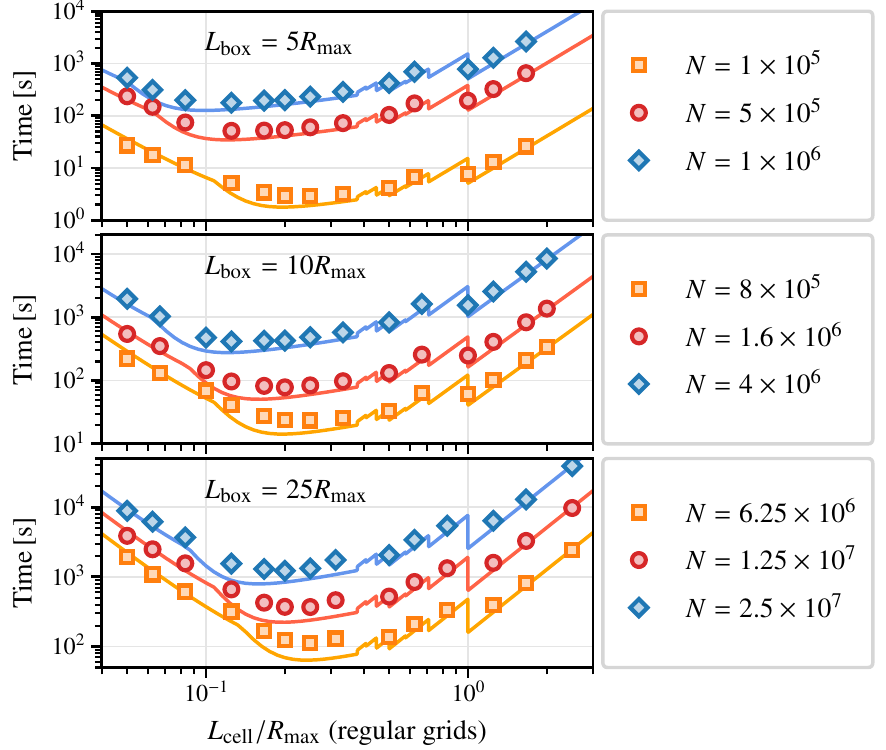}
\caption{Execution time of the grid-based pair counting routine with different cell sizes and query ranges, for periodic uniform random samples with different cubic box sizes and numbers of points. Solid lines show the best-fitting theoretical results detailed in Appendix~\ref{sec:complexity_struct}.}
\label{fig:grid_benchmark}
\end{figure}

We then perform a series of benchmarks with the pair counting routine based on regular grids, which reports simply the number of pairs with separations below $R_{\rm max}$, and omits histogram bins of distances to separate the impacts of the data structure and histogram update algorithm (see Sect.~\ref{sec:hist} for details).
For simplicity, the pair counting procedure is based on cubic grid cells with a side length of $L_{\rm cell}$, and run on $N$ uniformly distributed random points in a periodic cubic volume with the box size of $L_{\rm box}$.
The execution time of the pair counting processes with different settings are shown in Fig.~\ref{fig:grid_benchmark}, together with the theoretical model detailed in Appendix~\ref{sec:complexity_struct}.
Note that $L_{\rm box}$ and $L_{\rm cell}$ are expressed as factors of $R_{\rm max}$, as the benchmark results are irrelevant to the units of lengths.
The results confirm the sensitivity of computational costs to the cell sizes.
For the configurations we explore, the optimal $L_{\rm cell}$ is typically 0.1 to 0.5 times $R_{\rm max}$.

\subsection{\texorpdfstring{$k$}{k}-d tree}
\label{sec:struct_kd}

$k$-d tree \citep[][]{Bentley1975} is a binary space-partition data structure that is commonly used for range searching and nearest-neighbour search algorithms.
It partitions the $k$-dimensional space recursively with axis-aligned planes.
Depending on the choices of the splitting planes, there are several variants of the $k$-tree structure.
In this work we choose the \textit{optimised} $k$-d tree introduced by \citet[][]{Friedman1977}, for which the space-partition planes are perpendicular to the dimension with the largest data variance, and split the dataset into two parts at the median point.
Therefore, this variant always produces a balanced tree structure, and is particularly useful for observational catalogues with arbitrary survey geometry.

Algorithm~\ref{alg:kdtree} shows the procedure to construct the $k$-d tree for pair counting purposes.
The root node of the tree is associated with all the data points. For each non-leaf node, the two subsets of data after space partition are assigned to their two children, respectively.
In addition, we store the minimum AABB of points on each node for efficient data pruning, due to the simplicity of evaluating the minimum and maximum distances between AABBs, which can be good estimates of the separation ranges of points on different nodes.
Finally, the tree construction process is terminated when all leaf nodes contain at most $n_{\rm leaf}$ points.

\begin{algorithm}
\caption{\textsc{KdTree\_Build} ($\mathcal{P}$, $n_{\rm leaf}$)}\label{alg:kdtree}
\begin{algorithmic}[1]
\Require a point set $\mathcal{P}$ and the capacity of leaf nodes.
\Ensure the root of a $k$-d tree for $\mathcal{P}$.
\State Create a new node $\nu$, with $\nu.{\rm data} \gets \mathcal{P}$.
\State $\nu.{\rm bound} \gets$ \textsc{MinimumAABB} ($\mathcal{P}$)
\Comment{bounding volume of $\nu$}
\If{${\rm cardinality}(\mathcal{P}) \le n_{\rm leaf}$}
    \State \Return $\nu$ as a leaf node
\Else
    \State Find the axis direction with the largest coordinate variance for all points in $\mathcal{P}$, and divide $\mathcal{P}$ into $\mathcal{P}_1$ and $\mathcal{P}_2$ with a splitting plane perpendicular to this direction, such that $\mathcal{P}=\mathcal{P}_1 \cup \mathcal{P}_2$, $\mathcal{P}_1 \cap \mathcal{P}_2 = \varnothing$, and ${\rm cardinality}(\mathcal{P}_1) = \lfloor{\rm cardinality}(\mathcal{P}) / 2\rfloor$.
    \State $\nu.{\rm left} \gets$ \textsc{KdTree\_Build} ($\mathcal{P}_1$, $n_{\rm leaf}$)
    \Comment{left child of $\nu$}
    \State $\nu.{\rm right} \gets$ \textsc{KdTree\_Build} ($\mathcal{P}_2$, $n_{\rm leaf}$)
    \Comment{right child of $\nu$}
    \State \Return $\nu$
\EndIf
\end{algorithmic}
\end{algorithm}

Since the $k$-d tree is always balanced, there are in total $\mathcal{O}(N)$ tree nodes for a fixed $n_{\rm leaf}$.
The storage cost of the tree is then $\mathcal{O}(N)$.
Computations of the minimum AABB and coordinate variances require only two passes through the dataset.
Besides, we split the data for children nodes using the linear-time adaptive \textsc{QuickSelect} algorithm \citep[\texttt{MedianOfNinthers};][]{Alexandrescu2017}. Therefore, the $k$-d tree construction can be accomplished in $\mathcal{O} (N \log N)$ time, given the tree depth of $\mathcal{O}(\log N)$.

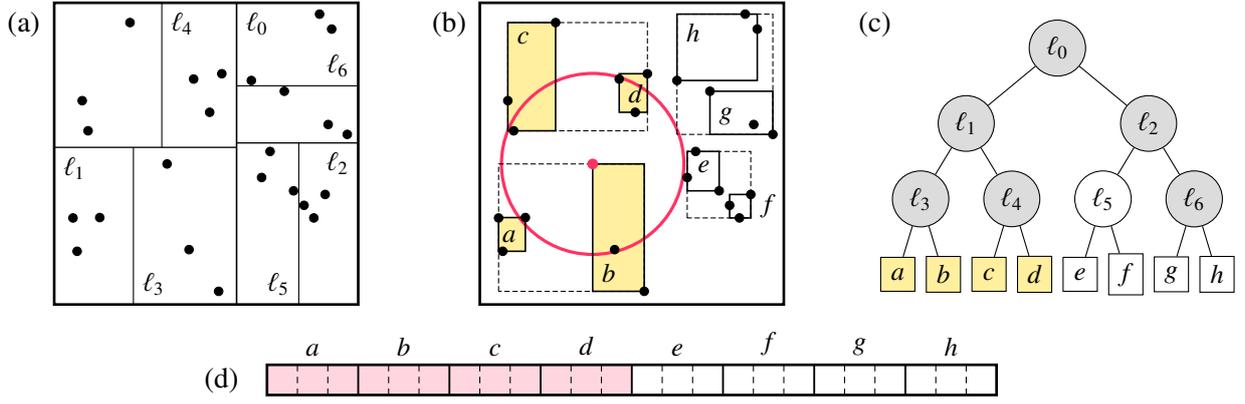
\begin{figure*}
\centering
\begin{tikzpicture}
\begin{scope}
\draw[split0] (5.9995,0) -- (5.9995,10) node[below right] {$\ell_0$};
\draw[split1] (0,5.2082) -- (5.9995,5.2082) node[pos=0, below right] {$\ell_1$};
\draw[split2] (2.6052,0) -- (2.6052,5.2082) node[pos=0, above right] {$\ell_3$};
\draw[split2] (3.5393,5.2082) -- (3.5393,10) node[below right] {$\ell_4$};
\draw[split1] (5.9995,5.3554) -- (10,5.3554) node[below left] {$\ell_2$};
\draw[split2] (8.0459,0) -- (8.0459,5.3554) node[pos=0, above left] {$\ell_5$};
\draw[split2] (5.9995,7.2469) -- (10,7.2469) node[above left] {$\ell_6$};
\draw plot[pointstyle] file {points.txt};
\draw[boundary] (0,0) rectangle (10,10);
\node[font=\large, align=right, below] at (-1,10) {(a)};
\end{scope}
\begin{scope}[xshift=14cm]
\fill[nodebgcolor] (0.618161,1.76381) rectangle (1.49798,2.88145);
\fill[nodebgcolor] (3.71239,0.436292) rectangle (5.40917,4.65917);
\fill[nodebgcolor] (0.921492,5.75724) rectangle (2.49305,9.34509);
\fill[nodebgcolor] (4.58547,6.37134) rectangle (5.51322,7.64931);
\draw[querypoint] (3.7123921,4.6591658) circle [radius=3];
\draw[parentnode] (0.618161,0.436292) rectangle (5.40917,4.65917);
\draw[parentnode] (0.921492,5.75724) rectangle (5.51322,9.34509);
\draw[parentnode] (6.82244,2.86932) rectangle (8.91325,5.07274);
\draw[parentnode] (6.48591,5.63799) rectangle (9.63929,9.62518);
\draw[leafnode] (0.618161,1.76381) rectangle (1.49798,2.88145);
\node[above left] at (1.49798,1.76381) {$a$};
\draw[leafnode] (3.71239,0.436292) rectangle (5.40917,4.65917);
\node[above right] at (3.71239,0.436292) {$b$};
\draw[leafnode] (0.921492,5.75724) rectangle (2.49305,9.34509);
\node[below right] at (0.921492,9.34509) {$c$};
\draw[leafnode] (4.58547,6.37134) rectangle (5.51322,7.64931);
\node[above right] at (4.58547,6.37134) {$d$};
\draw[leafnode] (6.82244,3.76765) rectangle (7.87392,5.07274);
\node[below left] at (7.87392,5.07274) {$e$};
\draw[leafnode] (8.21782,2.86932) rectangle (8.91325,3.64714);
\node[right] at (8.91325,3.2582) {$f$};
\draw[leafnode] (7.57011,5.63799) rectangle (9.63929,7.0684);
\node[above right] at (7.57011,5.63799) {$g$};
\draw[leafnode] (6.48591,7.42541) rectangle (9.12706,9.62518);
\node[below right] at (6.48591,9.62518) {$h$};
\draw plot[pointstyle] file {points.txt};
\filldraw[querypoint] (3.7123921,4.6591658) circle [radius=3.5pt];
\draw[boundary] (0,0) rectangle (10,10);
\node[font=\large, align=right, below] at (-1,10) {(b)};
\end{scope}
\begin{scope}[xshift=28cm]
\node[visitednode] at (5,8.5) {$\ell_0$}
    child{ node[visitednode] {$\ell_1$}
        child{ node[visitednode] {$\ell_3$}
            child{ node[visitedleaf] {$a$}}
            child{ node[visitedleaf] {$b$}}
        }
        child{ node[visitednode] {$\ell_4$}
            child{ node[visitedleaf] {$c$}}
            child{ node[visitedleaf] {$d$}}
        }
    }
    child{ node[visitednode] {$\ell_2$}
        child{ node[skippednode] {$\ell_5$}
            child{ node[skippedleaf] {$e$}}
            child{ node[skippedleaf] {$f$}}
        }
        child{ node[visitednode] {$\ell_6$}
            child{ node[skippedleaf] {$g$}}
            child{ node[skippedleaf] {$h$}}
        }
    }
;
\node[font=\large, align=right, below] at (-1,10) {(c)};
\end{scope}
\begin{scope}[xshift=7cm, yshift=-3cm]
\fill[visitedmem] (0,0) rectangle (12,1);
\draw[boundary] (0,0) rectangle (24,1);
\foreach \x in {0,...,7} {
    \draw[insideleaf] (\x*3+1,0) -- (\x*3+1,1);
    \draw[insideleaf] (\x*3+2,0) -- (\x*3+2,1);
    \draw[betweenleaf] (\x*3+3,0) -- (\x*3+3,1);
}
\foreach [count=\i] \j in {a,b,...,h}
    \node[above] at (\i*3-1.5, 1) {$\j$};
\node[font=\large, align=right] at (-1.5,0.5) {(d)};
\end{scope}
\end{tikzpicture}
\caption{Panels (a), (b) and (c) show partitions of the data points (black dots) during $k$-d tree construction, the resulting minimum axis-aligned bounding box of leaf nodes and their parents, as well as the diagram of the tree structure, respectively. In particular, non-leaf nodes in (c) are indicated by the corresponding dividing lines in (a). Red point and circle in panel (b) indicate the reference point and radius for a range searching. Grey regions in (c), as well as yellow areas in (b) and (c), highlight the visited non-leaf and leaf nodes respectively during this query.
The retrieved data points are shown in pink in panel (d).}
\label{fig:kdtree}
\end{figure*}

Fig.~\ref{fig:kdtree} shows the $k$-d tree constructed with $n_{\rm leaf} = 3$, upon the same sample points as in Fig.~\ref{fig:grid}. It can be seen that the space partition is adaptive, and in this particular example the tree is complete, with the same number of points on all leaf nodes.
In addition, the total AABB volume of nodes with the same depth can be significantly smaller than the volume of the full dataset, especially for the leaf nodes, due to the gaps between the bounding boxes of different nodes. This implies a relatively high data pruning efficiency, as it is easier to detect data groups that are too far away or too close to each other, compared to the grid-based method.
Actually, for the example shown in Fig.~\ref{fig:kdtree}, only four leaf nodes are visited after checking the distances between AABBs.
Besides, since the visited leaf nodes are in the same branch of the tree, the associate data points are continuously aligned in memory, which indicates a high memory access efficiency.

We use the dual-tree algorithm (see Sect.~\ref{sec:dual_tree}) for counting pairs with $k$-d tree, which traverses the tree nodes in a top-down manner. In brief, we skip all the descendants of two nodes when the separation range between the minimum AABBs of these nodes is entirely inside or outside the query range for pair counting. In other words, a leaf node is only visited if the corresponding AABB intersects with the query range boundary of its counterpart during the tree traversal process, which is usually a leaf node as well.
Consequently, the sizes of nodes from which the data points are retrieved are adaptive, and the number of visited nodes is greatly reduced compared to the grid-based approach, especially when the query range is large.

\begin{figure}
\centering
\includegraphics{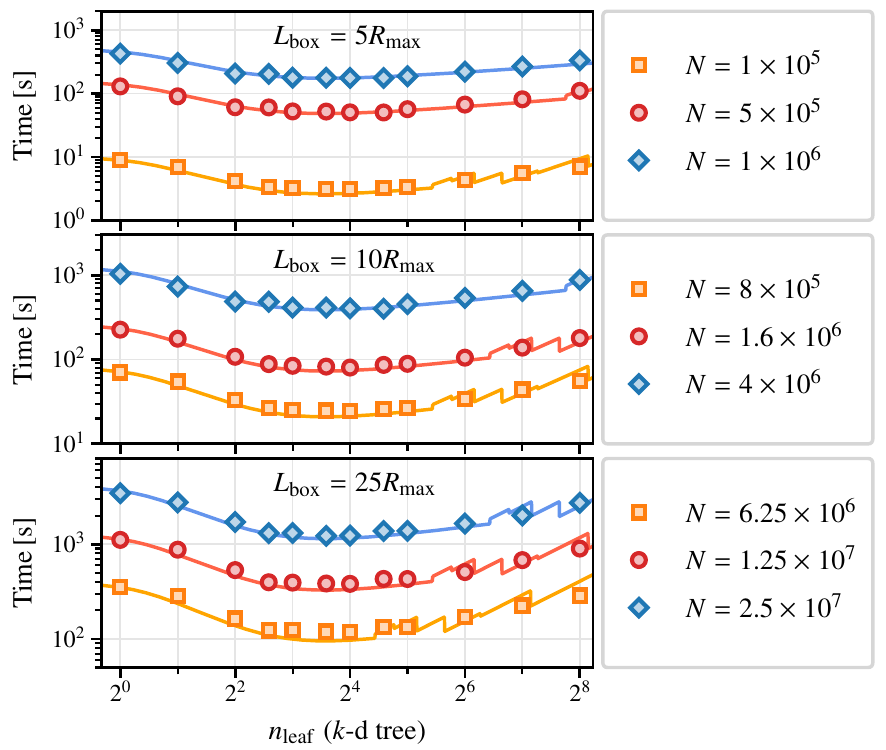}
\caption{Execution time of the pair counting routine based on $k$-d tree with different capacities of leaf nodes, for periodic uniform random samples with different cubic box sizes and numbers of points. Solid lines show the best-fitting theoretical results detailed in Appendix~\ref{sec:complexity_struct}.}
\label{fig:kdtree_benchmark}
\end{figure}

We then run the pair counting routine based on $k$-d tree, upon the same catalogues used for benchmarks of the grid-based method.
Again, we consider a single histogram bin for separations below $R_{\rm max}$.
The results with different choices of leaf node capacity are shown in Fig.~\ref{fig:kdtree_benchmark}.
It turns out that the execution time of the pair counting algorithm based on $k$-d tree does not vary significantly as $n_{\rm leaf}$, especially when $4 \le n_{\rm leaf} \le 64$, compared to the strong cell-size dependence of the grid-based method (see Fig.~\ref{fig:grid_benchmark}, and more discussions are detailed in Appendix~\ref{sec:complexity_struct}).
Moreover, the optimal $n_{\rm leaf}$ is found to be 8 for almost all configurations studied in this work.
This makes the $k$-d tree structure particularly useful in practice, as it is not necessary to explore different choices of $n_{\rm leaf}$ to maximise the pair counting efficiency for different input samples.

\subsection{Ball tree}
\label{sec:struct_ball}

Similar to $k$-d tree, ball tree \citep[][]{Omohundro1989} is also a binary space partition tree that is useful for range queries, especially for high dimensions.
In general, every node of a ball tree defines a hypersphere, that contains all the points on the node.
This makes it slightly easier to compute the minimum and maximum distances between two nodes, compared to the case of axis-aligned boxes for the $k$-d tree.
However, for traditional ball tree implementations \citep[e.g.][]{Moore2000}, the tree is not necessarily balanced, and the hyperspheres, or balls, can be significantly larger than the minimum bounding spheres of the points.
As a result, both the dual-tree algorithm (see Sect.~\ref{sec:dual_tree}) and the data pruning process are sub-optimal for pair counting \citep[cf. however an application in][]{correlcalc}.
We then introduce a new variant of the ball tree structure to circumvent these problems.

To construct a balanced ball tree, one way is to use the space partition scheme of the $k$-d tree. In this case, all the subsets of data points are bounded by axis-aligned boxes, and the data pruning with minimum bounding spheres is supposed to be less efficient than that of the $k$-d tree, due to their generally larger volumes than the corresponding minimum AABBs.
Moreover, axis-aligned partition schemes may be sub-optimal for observational data with complicated shapes.
To circumvent these problems, we follow the space partition approach introduced by \citet[][]{Dolatshah2015}, which defines the splitting plane based on the principal component analysis (PCA).
In particular, the plane is chosen to be perpendicular to the most significant principal component of the data distribution, which is the direction with the largest variance of the data points.
Thus, the resulting subsets of data are statistically the least extended. In this way, the minimum bounding spheres of the ball tree nodes are generally small enough in practice, for efficient data pruning.

The next step is to compute the minimum bounding spheres of the subdivided datasets. In principle, the exact solution can be obtained in linear time using a randomised algorithm \citep[][]{Welzl1991,Gartner1999}. However, it is relatively slow for a large dataset. We then focus on the approximate algorithm introduced by \citet[][]{Ritter1990}, which ensures that all the input data points are enclosed by the reported sphere, but typically overestimates the radius by $\lesssim 20$ per cent \citep[e.g.][]{Larsson2008}.
This algorithm set up an initial sphere with three points that are far away from each other, and then go through the rest of the data points. Whenever a point is found outside the sphere, a new sphere that encloses both the point and the previous sphere is constructed.
Following the spirit of \citet[][]{Larsson2008}, we improve this algorithm by constructing a better initial sphere, which is defined by the extreme points along the directions of the first two principal components. In practice, the minimum bounding sphere of the four extreme points is computed exactly, and this sphere is updated in the same way as in \citet[][]{Ritter1990}.

The full procedure for the construction of our ball tree variant is shown in Algorithm~\ref{alg:balltree}, which is very similar to that of the $k$-d tree (see Algorithm~\ref{alg:kdtree}), and consumes $\mathcal{O} (N)$ space as well since the tree is balanced.
In practice, we rely on the symmetric QR algorithm \citep[e.g.][]{Golub2013} for the 3D PCA.
When the number of data points is large, the computing time for PCA is dominated by the covariance matrix evaluation, which requires two passes through the dataset. The update of the minimum bounding sphere needs another pass. Again, we use the adaptive \textsc{QuickSelect} algorithm for the data partition, but with a comparison rule that involves the first principal component of the dataset. Therefore, the time complexity for construction a single ball tree node is $\mathcal{O}(N)$, and it is $\mathcal{O} (N \log N)$ for the full tree. In practice, the ball tree construction process is typically only marginally slower than that of the $k$-d tree.

\begin{algorithm}
\caption{\textsc{BallTree\_Build} ($\mathcal{P}$, $n_{\rm leaf}$)}\label{alg:balltree}
\begin{algorithmic}[1]
\Require a point set $\mathcal{P}$ and the capacity of leaf nodes.
\Ensure the root of a ball tree for $\mathcal{P}$.
\State Create a new node $\nu$, with $\nu.{\rm data} \gets \mathcal{P}$.
\State Compute $\boldsymbol{u}_1$ and $ \boldsymbol{u}_2$, the first two principal components of $\mathcal{P}$.
\State $\mathcal{E} \gets$ \textsc{FindExtremePoints} ($\mathcal{P}$, $\{ \boldsymbol{u}_1, \boldsymbol{u}_2 \}$)
\State $B \gets$ \textsc{MinimumBoundingSphere} ($\mathcal{E}$)
\ForAll{$\boldsymbol{p} \in \mathcal{P} \setminus \mathcal{E}$}
    \IIf{$\boldsymbol{p}$ outside $B$}
        $B \gets$ \textsc{GrowSphere} ($B$, $\boldsymbol{p}$)
    \EndIIf
\EndFor
\State $\nu.{\rm bound} \gets B$ \Comment{bounding volume of $\nu$}
\If{${\rm cardinality}(\mathcal{P}) \le n_{\rm leaf}$}
    \State \Return $\nu$ as a leaf node
\Else
    \State Divide $\mathcal{P}$ into subsets $\mathcal{P}_1$ and $\mathcal{P}_2$, such that $\mathcal{P}=\mathcal{P}_1 \cup \mathcal{P}_2$, $\mathcal{P}_1 \cap \mathcal{P}_2 = \varnothing$, $\max_{(\boldsymbol{p}_1 \in \mathcal{P}_1)} ( \boldsymbol{p}_1 \cdot \boldsymbol{u}_1 ) \le \min_{(\boldsymbol{p}_2 \in \mathcal{P}_2)} ( \boldsymbol{p}_2 \cdot \boldsymbol{u}_1 )$, and ${\rm cardinality}(\mathcal{P}_1) = \lfloor{\rm cardinality}(\mathcal{P}) / 2\rfloor$.
    \State $\nu.{\rm left} \gets$ \textsc{BallTree\_Build} ($\mathcal{P}_1$, $n_{\rm leaf}$)
    \Comment{left child of $\nu$}
    \State $\nu.{\rm right} \gets$ \textsc{BallTree\_Build} ($\mathcal{P}_2$, $n_{\rm leaf}$)
    \Comment{right child of $\nu$}
    \State \Return $\nu$

\EndIf
\end{algorithmic}
\end{algorithm}

\begin{figure*}
\centering
\begin{tikzpicture}
\begin{scope}
\draw[split0] (8.70156,0) -- (4.42839,10) node[pos=0, above left] {$\ell_0$};
\draw[split1] (0,3.86849) -- (6.73914,4.59242) node[pos=0, above right] {$\ell_1$};
\draw[split2] (2.52711,4.13996) -- (3.70563,10) node[below left] {$\ell_3$};
\draw[split2] (3.05864,0) -- (2.89485,4.17946) node[pos=0, above left] {$\ell_4$};
\draw[split1] (6.35707,5.48654) -- (10,5.87378) node[above left] {$\ell_2$};
\draw[split2] (5.99121,10) -- (8.79971,5.74619) node[pos=0, below] {$\ell_5$};
\draw[split2] (6.95651,4.08374) -- (10,3.28671) node[pos=0.84, below] {$\ell_6$};
\draw plot[pointstyle] file {points.txt};
\draw[boundary] (0,0) rectangle (10,10);
\node[font=\large, align=right, below] at (-1,10) {(a)};
\end{scope}
\begin{scope}[xshift=14cm]
\fill[nodebgcolor] (1.80363,7.55117) circle[radius=1.92184] node {$a$};
\fill[nodebgcolor] (4.14893,6.06613) circle[radius=1.47313] node {$b$};
\begin{scope} 
\clip (1.80363,7.55117) circle[radius=1.92184];
\fill[nodebgcolor2] (4.14893,6.06613) circle[radius=1.47313];
\end{scope}
\fill[nodebgcolor] (1.06575,2.36277) circle[radius=0.675173] node {$c$};
\fill[nodebgcolor] (6.11581,2.32289) circle[radius=2.01459] node {$d$};
\fill[nodebgcolor] (6.54166,7.35885) circle[radius=1.06867] node[above left] {$e$};
\draw[querypoint] (3.7123921,4.6591658) circle [radius=3];
\begin{scope}
\clip(0,0) rectangle (10,10); 
\draw[parentnode] (3.10272,7.00213) circle[radius=2.42098];
\draw[parentnode] (3.78828,2.98665) circle[radius=3.27131];
\draw[parentnode] (7.69394,7.70496) circle[radius=2.18143];
\draw[parentnode] (8.56944,4.4602) circle[radius=1.59115];
\draw[leafnode] (1.80363,7.55117) circle[radius=1.92184] node {$a$};
\draw[leafnode] (4.14893,6.06613) circle[radius=1.47313] node {$b$};
\draw[leafnode] (1.06575,2.36277) circle[radius=0.675173] node {$c$};
\draw[leafnode] (6.11581,2.32289) circle[radius=2.01459] node {$d$};
\draw[leafnode] (6.54166,7.35885) circle[radius=1.06867] node[above left] {$e$};
\draw[leafnode] (8.87007,7.79565) circle[radius=1.83497] node {$f$};
\draw[leafnode] (8.20703,3.31848) circle[radius=0.559209];
\node[below right] at (8.4,3) {$g$};
\draw[leafnode] (8.46276,4.93924) circle[radius=1.36838] node {$h$};
\end{scope}
\draw plot[pointstyle] file {points.txt};
\filldraw[querypoint] (3.7123921,4.6591658) circle [radius=3.5pt];
\draw[boundary] (0,0) rectangle (10,10);
\node[font=\large, align=right, below] at (-1,10) {(b)};
\end{scope}
\begin{scope}[xshift=28cm]
\node[visitednode] at (5,8.5) {$\ell_0$}
    child{ node[visitednode] {$\ell_1$}
        child{ node[visitednode] {$\ell_3$}
            child{ node[visitedleaf] {$a$}}
            child{ node[visitedleaf] {$b$}}
        }
        child{ node[visitednode] {$\ell_4$}
            child{ node[visitedleaf] {$c$}}
            child{ node[visitedleaf] {$d$}}
        }
    }
    child{ node[visitednode] {$\ell_2$}
        child{ node[visitednode] {$\ell_5$}
            child{ node[visitedleaf] {$e$}}
            child{ node[skippedleaf] {$f$}}
        }
        child{ node[skippednode] {$\ell_6$}
            child{ node[skippedleaf] {$g$}}
            child{ node[skippedleaf] {$h$}}
        }
    }
;
\node[font=\large, align=right, below] at (-1,10) {(c)};
\end{scope}
\begin{scope}[xshift=7cm, yshift=-3cm]
\fill[visitedmem] (0,0) rectangle (15,1);
\draw[boundary] (0,0) rectangle (24,1);
\foreach \x in {0,...,7} {
    \draw[insideleaf] (\x*3+1,0) -- (\x*3+1,1);
    \draw[insideleaf] (\x*3+2,0) -- (\x*3+2,1);
    \draw[betweenleaf] (\x*3+3,0) -- (\x*3+3,1);
}
\foreach [count=\i] \j in {a,b,...,h}
    \node[above] at (\i*3-1.5, 1) {$\j$};
\node[font=\large, align=right] at (-1.5,0.5) {(d)};
\end{scope}
\end{tikzpicture}
\caption{Same as Fig.~\ref{fig:kdtree}, but for a variant of ball tree. Panel (a) shows the partition lines based on the principle component analysis, and panel (b) shows the resulting minimum bounding spheres of leaf nodes and their parents.}
\label{fig:balltree}
\end{figure*}
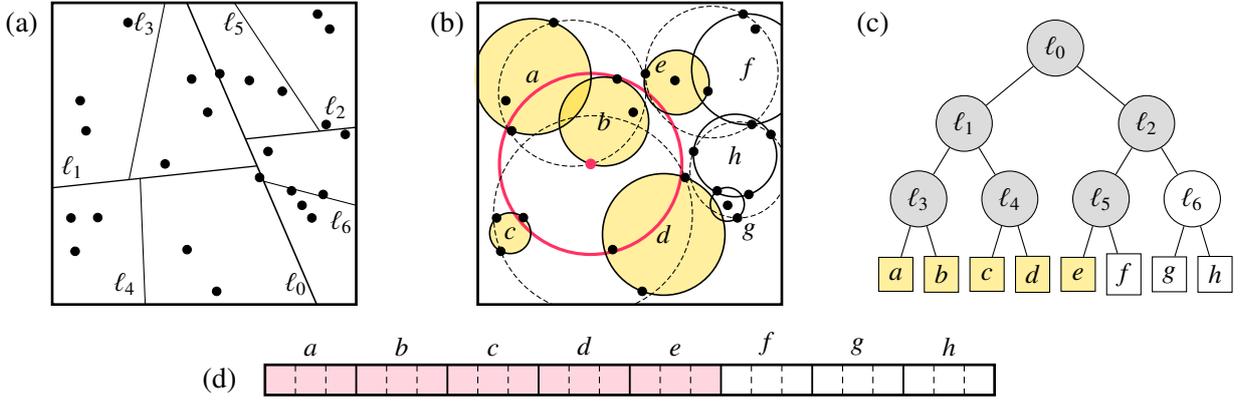

An example of the ball tree constructed with the points as in Fig.~\ref{fig:grid} is shown in Fig.~\ref{fig:balltree}. The space partition lines are not axis-aligned in general, resulting in different data point groups than those of the axis-aligned partition scheme (see Fig.\ref{fig:kdtree}).
Note that different nodes with the same depth do not share data points, albeit their bounding spheres may overlap.
The bounding sphere of a node may not be fully inside that of its parent. This does not necessarily mean a low data pruning efficiency, as the distances between different nodes are always examined in the top-down order.
For the example in Fig.~\ref{fig:balltree}, the range searching involves one more leaf node than that of the $k$-d tree, but the visited data points are still stored continuously, indicating a good memory locality.

\begin{figure}
\centering
\includegraphics{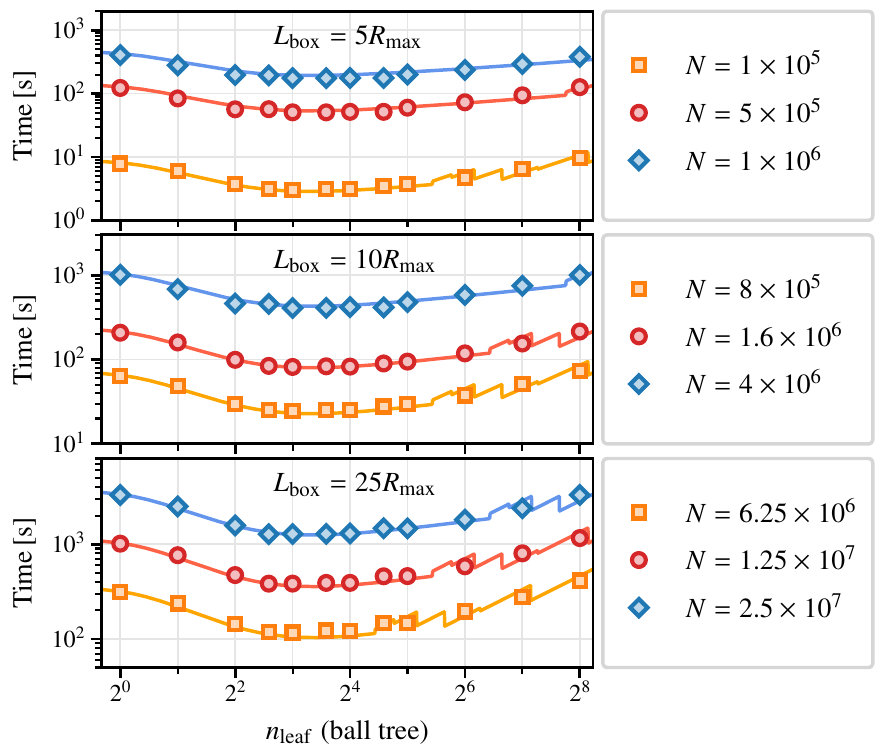}
\caption{Execution time of the pair counting routine based on ball tree with different capacities of leaf nodes, for periodic uniform random samples with different cubic box sizes and numbers of points. Solid lines show the best-fitting theoretical results detailed in Appendix~\ref{sec:complexity_struct}.}
\label{fig:balltree_benchmark}
\end{figure}

Similar to the case of $k$-d tree, the tree-independent dual-tree algorithm (see Sect.~\ref{sec:dual_tree}) is used for counting pairs with ball tree.
The benchmark results with ball tree are shown in Fig.~\ref{fig:balltree_benchmark}, with a single histogram bin for separations below $R_{\rm max}$.
One can see that the dependences of execution time measurements on $n_{\rm leaf}$ are similar to those of the $k$-d tree.
Actually, the theoretical model is derived for $k$-d tree (see Appendix~\ref{sec:complexity_struct}), but turns out to work well for ball tree as well.
This can be explained by the fact that the spatial partition schemes are similar for these two data structures for a periodic box.
Again, the results are not sensitive to the choice of $n_{\rm leaf}$, and a leaf node capacity of 8 is found to be optimal for almost all cases.

\subsection{Comparisons of the data structures}
\label{sec:compare_struct}

In order to identify the optimal data structure among the ones discussed so far for real-world pair counting problems, we perform two additional sets of benchmarks with both periodic and non-periodic datasets.
For tests with periodic boundary condition, which is the case for cosmological simulations, we generate uniformly distributed random points in a cubic volume, with the box size of $L_{\rm box}$.
Then, to mimic the geometry of the observational data in redshift bins, we cut the cubic catalogues at $R_{\rm out} = L_{\rm box}$ and $R_{\rm in} = L_{\rm box} / 2$ with respect to a corner of the boxes, and take the sections in between as our non-periodic samples, which are essentially octants of spherical shells.
Again, we count pairs in $[0,R_{\rm max})$ as a whole to exclude costs of the histogram update process.
In particular, for all tests we set $L_{\rm box} = 10 R_{\rm max}$, which is typical for modern cosmological applications, e.g., pair counting with separations up to $200\,h^{-1}\,{\rm Mpc}$, for simulations with the side length of $2\,h^{-1}\,{\rm Gpc}$.
We consider only cubic grid cells for regular grids, but with two choices of cell sizes, $0.1 R_{\rm max}$ and $0.2 R_{\rm max}$, which are near optimum for most cases shown in Fig.~\ref{fig:grid_benchmark}.
Meanwhile, we set $n_{\rm leaf} = 8$ for both $k$-d and ball trees as it is shown to be the most favourable for almost all cases in Figs.~\ref{fig:kdtree_benchmark} and \ref{fig:balltree_benchmark}.

\begin{figure}
\centering
\includegraphics{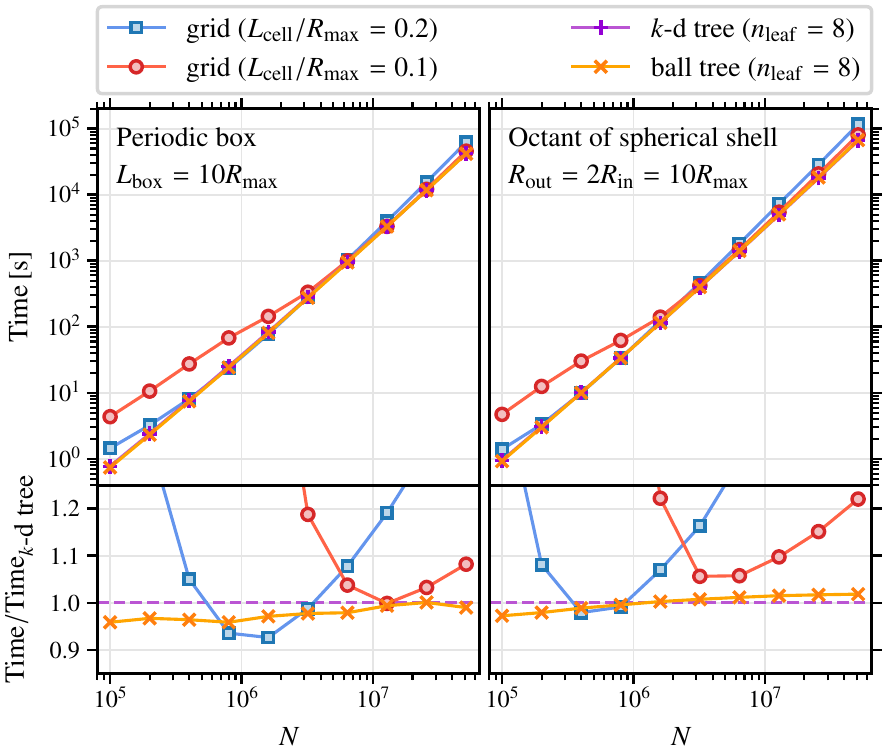}
\caption{Comparisons of the computational costs of pair counting routines based on different data structures, for periodic uniform random samples with different numbers of points in cubic volume (\textit{left}), and sections between $R_{\rm in}$ and $R_{\rm out}$ of the same catalogues with respect to a corner of the boxes, where $R_{\rm out}$ is equal to the box size (\textit{right}).}
\label{fig:struct_compare}
\end{figure}

The benchmark results with different input sample sizes are shown in Fig.~\ref{fig:struct_compare}.
It can be seen that the performances of the two tree structures are very close, with the differences $<5$ per cent for all our tests.
Besides, we find again that the efficiency of the grid-based method is sensitive to the choice of cell size.
In particular, the optimal cell size decreases as the increase of the sample size.
When the number of data points is $\lesssim 10^6$, regular grids with the optimal cell size can be slightly better than the trees, but the improvement is only $\lesssim 5$ per cent compared to the ball tree.
However, if a sub-optimal cell size is used, the computing time with regular grids can be as large as twice that of the trees.
When the number of data points is $\gtrsim 10^7$, the tree structures are always favoured in terms of computational costs, regardless of the choice of cell size for regular grids, especially for the non-periodic and non-cubic catalogues.

To conclude, $k$-d tree and ball tree both perform superior to regular grids for modern and next-generation cosmological pair counting problems with $\gtrsim 10^7$ objects in the data or random catalogues, due to the lower computational costs in general, as well as the absence of fine-tuning parameters that depend on the input samples and strongly affect the performances.
Since there is no essential difference in the efficiencies of the two tree structures, we implement both $k$-d tree and ball tree in the \fcfc{} toolkit.

\subsection{Discussions on data structures for pair counting}
\label{sec:discuss_struct}

There are a variety of data structures for different range searching problems in the field of computational geometry \citep[e.g.][]{deBerg2008}.
The basic idea of the time-efficient data structures is to allow the report of groups of points directly, without visiting them individually.
Therefore, in principle the complexity of a pair counting algorithm can be better than $\mathcal{O} (N^2)$ -- the complexity of a brute-force approach which examines all data pairs.
However, for cosmological applications, it is generally necessary to count pairs in thin separation bins.
In fact, the 2PCFs are typically measured in $(s, \mu)$ or $(\sigma, \pi)$ bins for anisotropic information, which are given by
\begin{align}
&s = | \bm{s} | = | \bm{s}_2 - \bm{s}_1 | , \\
&\pi = \frac{| \bm{s} \cdot \bm{l} |}{| \bm{l} |} , \\
&\sigma = \sqrt{s^2 - \pi^2} , \\
&\mu = \pi / s ,
\end{align}
where $\bm{s}_1$ and $\bm{s}_2$ denote the coordinates of two points forming a pair, and $\bm{l}$ is the line-of-sight vector.
For observational data, $\bm{l}$ is typically defined as
\begin{equation}
\bm{l}_{\rm obs} = \bm{s}_2 + \bm{s}_1 ,
\end{equation}
while for simulations the plane-parallel line-of-sight is usually assumed, e.g.,
\begin{equation}
\bm{l}_{\rm sim} = \hat{\bm{e}}_z = (0,0,1) .
\end{equation}

The complicated binning schemes make it difficult to find pairs of large data groups with separations all in the same bin.
For instance, the typical number density of modern galaxy samples is $\rho \sim 10^{-3}\,h^3\,{\rm Mpc}^{-3}$.
Then there are on average only one point in a cubic volume with the box size of $10\,h^{-1}\,{\rm Mpc}$, which is already larger than the commonly used separation bin width of $5\,h^{-1}\,{\rm Mpc}$ for 2PCFs.
This problem may be less severe for galaxy catalogues with strong clustering patterns. But the most challenging tasks are normally pair counting with random samples for the normalisation of 2PCFs.
Thus, in most cases one has to visit individual pairs for updating pair counting histograms.

For a 3D periodic box, when $\rho R_{\rm max}^3 \gg 1$, the total number of pairs with separations in $[0,R_{\rm max})$ can be estimated by
\begin{equation}
\hat{N}_{\rm pair} = N \cdot \rho \frac{4\uppi R_{\rm max}^3}{3} = N^2 \cdot \frac{4\uppi}{3} \left( \frac{R_{\rm max}}{L_{\rm box}} \right)^3 .
\label{eq:npair_valid}
\end{equation}
Since $\hat{N}_{\rm pair} \propto N^2$, the complexity of a real-world pair counting algorithm is generally ineluctably $\mathcal{O} (N^2)$.
For this reason, the aim of the data structures described in this work is to reduce the constant factor hidden in the complexity by efficient data pruning. After all, $\hat{N}_{\rm pair} / N^2$ can be as small as $\sim 10^{-3}$ when $L_{\rm box} = 10 R_{\rm max}$.
Hence the pair counting algorithm with a well designed data structure can still be faster than the brute-force approach by a few orders of magnitude.

Note however that it is in principle possible to reduce the complexity of the pair counting process for certain cosmological problems.
As an example, since $\sigma$ and $\pi$ are independent with the plane-parallel line-of-sight, the evaluation of $(\sigma,\pi)$ pair counts for periodic simulations can be benefited from developments of orthogonal range queries \citep[][]{deBerg2008}.
For instance, following the spirit of the range tree \citep[e.g.][]{Lueker1978}, one can construct a binary tree with the $z$ coordinates, and on each node there can be an associate $k$-d tree or range tree for the $x$ and $y$ coordinates.
Then, groups of pairs in $\pi$ bins can be reported in logarithmic time, and individual pair visits are only required for the associate 2D subtrees. This improves the overall pair counting complexity with additional storage space. We leave detailed studies on this case to a future work.

For future samples with unprecedented number densities, isotropic pair counting with $s$ bins only can potentially be improved as well.
For a given reference point $(x_0, y_0, z_0)$, the pair counting process is equivalent to a spherical range searching, i.e., finding all points $(x,y,z)$ within a certain radius $R$:
\begin{equation}
(x - x_0 )^2 + (y - y_0 )^2 + (z - z_0 )^2 < R^2 .
\end{equation}
Defining $w \equiv x^2 + y^2 + z^2$, the condition can be rewritten as
\begin{equation}
2 x_0 x + 2 y_0 y + 2 z_0 z - w - w_0 + R^2 > 0 .
\end{equation}
Therefore, the 3D spherical range searching problem is converted to a 4D half-space range search, i.e., finding all the points $(x,y,z,w)$ above a given hyperplane.
This is a well known problem in computational geometry, and there exists data structures that are able to accomplish the query in logarithmic time.
Tradeoffs between the query time and storage costs are also possible \citep[see][for reviews]{deBerg2008,Agarwal2017}.
But the data structures and algorithms are generally very difficult to implement in practice. We leave them for future developments.

It is possible to further boost hugely the efficiency of pair counting by allowing inexact solutions.
For instance, there are data structures for approximate range queries with controlled errors, which can be adjusted to vary the query time and storage costs \citep[e.g.][]{Guilherme2020}.
There are also 2PCF estimators that pixelate the volume \citep[][]{CUTE}, neglect the extents of tree nodes that are far away from each other \citep[][]{Zhang2005}, or make use of Fast Fourier Transforms \citep[e.g.][]{Pen2003}.
It is important to validate these approximate methods in terms of the accuracies on different scales with modern cosmological data.
We will perform relevant tests and combine both exact and inexact methods in \fcfc{} to achieve higher efficiency with tuneable precision in a following paper.


\section{Algorithms}
\label{sec:alg}

Algorithms are another fundamental building block of a program besides data structures.
A good algorithm may accomplish computational tasks efficiently by taking advantage of the layout of input datasets in memory given the data structure, or making use of memoization, which avoids redundant computations.
The most important algorithms used by \fcfc{} are the ones for identifying pairs within desired separation ranges, and updating histogram bins given a large number of (multi-dimensional) pair separations, which are usually the most time consuming tasks for a correlation function calculator.

\subsection{Tree-independent dual-tree algorithm}
\label{sec:dual_tree}

With the tree structures described in the previous section, one can avoid a considerable fraction of unnecessary distance evaluations, provided an algorithm which detects node pairs that are not in the separation range of interest as early as possible.
To this end, it is preferred to traverse trees in a top-down manner, since if the separation range between a pair of parent nodes is entirely outside or inside the query range, all their descendant nodes can be omitted.
In particular, for the latter case, we visit the data associated with the parent nodes directly, to avoid unnecessary tree node visits.
We then end up with Algorithm~\ref{alg:dual_tree}, which is an improved version of the dual-tree algorithm introduced by \citet[][]{Moore2001}.
Note that our dual-tree algorithm is tree-independent \citep[see also][]{Curtin2013}. Therefore, it is applicable to all binary space-partition tree structures in principle.
The complexity of the dual-tree algorithm should depend on the tree structure, though in practice the pair counting efficiencies with the $k$-d and ball trees are quite similar (see Fig.~\ref{fig:struct_compare}).

\begin{algorithm}
\caption{\textsc{PairCount\_DualTree} ($\mathcal{N}$, $\mathcal{S}$, $\mathcal{H}$)}\label{alg:dual_tree}
\begin{algorithmic}[1]
\Require a stack $\mathcal{N}$ for pairs of tree nodes, the separation range $\mathcal{S}$ of interest, and the histogram $\mathcal{H}$ for storing pair counts.
\State Pop a pair of tree nodes $\{ \nu_1 , \nu_2 \}$ from $\mathcal{N}$.
\If{\textsc{DistanceRange} ($\nu_1.{\rm bound}$ , $\nu_2.{\rm bound}$) $\cap \,\mathcal{S} = \varnothing$}
    \State \Return \Comment{descendants of both nodes are pruned}
\ElsIf{\textsc{DistanceRange} ($\nu_1.{\rm bound}$ , $\nu_2.{\rm bound}$) $\subseteq \mathcal{S}$ \OR $\nu_1$ and $\nu_2$ are both leaves}
	\ForAll{$\boldsymbol{p}_1 \in \nu_1.{\rm data}$, $\boldsymbol{p}_2 \in \nu_2.{\rm data}$}
        \State $d \gets$ \textsc{Distance} ($\boldsymbol{p}_1$, $\boldsymbol{p}_2$)
        \IIf{$d \in \mathcal{S}$}
            update histogram $\mathcal{H}$ with $d$
        \EndIIf
    \EndFor
\ElsIf{neither of $\nu_1$ and $\nu_2$ is a leaf node}
    \State Push $\{ \nu_1.{\rm right}, \nu_2.{\rm right} \}$ and $\{ \nu_1.{\rm right}, \nu_2.{\rm left} \}$ onto $\mathcal{N}$.
    \State Push $\{ \nu_1.{\rm left}, \nu_2.{\rm right} \}$ and $\{ \nu_1.{\rm left}, \nu_2.{\rm left} \}$ onto $\mathcal{N}$.
\ElsIf{$\nu_1$ is a leaf}
    \State Push $\{ \nu_1, \nu_2.{\rm right} \}$ and $\{ \nu_1, \nu_2.{\rm left} \}$ onto $\mathcal{N}$.
\Else \Comment{$\nu_2$ is a leaf, but $\nu_1$ is not}
    \State Push $\{ \nu_1.{\rm right}, \nu_2 \}$ and $\{ \nu_1.{\rm left}, \nu_2 \}$ onto $\mathcal{N}$.
\EndIf
\end{algorithmic}
\end{algorithm}

Our algorithm traverses the tree in the so-called depth-first order, as it uses less memory than the breadth-first order for a balanced tree.
This is because at a given level, the depth of the balanced binary tree is generally smaller than the width.
We then maintain a stack for pairs of tree nodes, to avoid recursive function calls in typical depth-first dual-tree algorithms \citep[][]{Moore2001,March2012}.
This increases the scalability of the algorithm with parallelisation, as different threads are able to work independently given their private stacks for dual nodes (see Sect.~\ref{sec:openmp}). The overhead due to the stack memory cost for recursive function calls is also mitigated.
Note that we do not directly report the total number of pairs from two nodes, as is done in \citet[][]{Moore2001}, since the examination of individual pairs is usually necessary for histogram updates with separation bins (see Sect.~\ref{sec:discuss_struct} for details).

Though not shown explicitly, the implementation of Algorithm~\ref{alg:dual_tree} in \fcfc{} is further optimised for some specific but common cases.
For auto pair counts we discard node pairs $\{ \nu_2 , \nu_1 \}$ when $\{ \nu_1 , \nu_2 \}$ is (going to be) visited, to avoid duplicate pair examinations, as $\nu_1$ and $\nu_2$ belong to the same tree.
Besides, following \citet[][]{CORRFUNC}, we do not inspect individual pairs of data points for wrapping large separations when periodic boundary conditions are enabled.
Instead, we compute the offsets of coordinates for the periodic wrapping of node pairs given their bounding volumes, and apply the offsets directly to all the associate data points. In this way, a large number of periodic boundary detections are avoided.

Note also that the dual-tree algorithm can be applied to angular pair counts directly, and can be easily extended for higher order statistics, such as 3- or 4-point correlation functions.
We leave relevant developments to future work.

\subsection{Update of pair counting histograms}
\label{sec:hist}

The cost of histogram updates in Algorithm~\ref{alg:dual_tree} can be considerable, as there are usually numerous pairs within the query range, that scales with $\mathcal{O} (N^2)$ for most cases (see Eq.~\eqref{eq:npair_valid}).
Therefore, the complexity of the histogram update process is usually $\mathcal{O} (N^2)$, which is independent of data structures and algorithms.
Nevertheless, it is possible to reduce the hidden constant factor with a smart algorithm. In general, this factor relies on the number of bins and the distribution of separations, which are then crucial for comparing the performances of different histogram update algorithms.
In practice, pair separations are usually computed from the squared distances. Hence we sample squared distances randomly following their expected distributions with a periodic box (see Appendix~\ref{sec:dist_s2} for details), for the histogram update algorithm benchmarks.
We assume monotonically increasing histogram bins for the tests. In reality this can be fulfilled by pre-sorting the bins. We also require the bins to be continuous, which is a common scenario in practice. Furthermore, we use zero-based bin indices throughout this work.

\subsubsection{Comparison-based methods}
\label{sec:hist_comp}

A direct way of locating the histogram bins of given separation values is to compare them with the bin edges. In this case, the squared distances can be compared against pre-computed squared bin edges, without evaluating square roots for the actual separations.
This improves both the efficiency and numerical stability of the algorithms.
A commonly used method for this purpose is the binary search algorithm.
The average complexity of this algorithm is $\mathcal{O} (\log N_{\rm bin})$, where $N_{\rm bin}$ denotes the number of histogram bins.
This complexity is optimal for comparison-based methods when the separations are distributed uniformly across the bins and come in random order.

However, in reality there are usually more pairs with larger separations (see Appendix~\ref{sec:dist_s2}). Thus, it is worthwhile to consider a simple algorithm that traverse histogram bins continuously in the reverse order, i.e., starting from the bin for the largest separations \citep[see][]{CORRFUNC}.
The worst-case complexity of this algorithm is $\mathcal{O} (N_{\rm bin})$.
Nevertheless, the average computational cost can be smaller than that of the binary search algorithm, especially when the distribution of separations across the bins is highly asymmetric.

In principle, comparison-based methods can be further improved by taking advantage of the locality of separation values during the pair counting process.
This is particularly true for the tree structures discussed in Sect.~\ref{sec:data_struct}, which group nearby data points together.
In this case, splay tree is a potentially useful data structure for histogram updates, with which frequently accessed bins can be visited more quickly \citep[][]{splaytree}.
Nevertheless, the performances of comparison-based methods are limited by the $N_{\rm bin}$ dependences and unavoidable conditional branches, which are harmful to the performance of instruction-level parallelism with modern pipelined processors.
Given also the high efficiency of alternative algorithms introduced later, we do not implement splay tree in this work.

\subsubsection{Index mapping functions}
\label{sec:hist_map_func}

The pair separation histogram can be updated in constant time and without branches if it is possible to map squared distances directly onto indices of the corresponding histogram bins.
For evenly spaced bins on both linear and logarithmic scales, which are the most common configurations in practice, the index mapping forms are simple.
Thus, it is of practical interest to examine index mapping algorithms for these specific cases.

For uniform linear separation bins in the range of $[s_{\rm min}, s_{\rm max})$, the index of the bin for a given squared distance $s^2$ is
\begin{equation}
\mathfrak{i}_{\rm lin} (s^2) = \left\lfloor \frac{ \sqrt{s^2} - s_{\rm min} }{ \Delta_{\rm lin} s  } \right\rfloor ,
\quad s_{\rm min}^2 \le s^2 < s_{\rm max}^2 ,
\label{eq:idx_map_lin}
\end{equation}
where $\Delta_{\rm lin} s$ indicates the width of the bins.
Given the fact that the pair counting process is independent of coordinate units, we can rescale data point coordinates and histogram bins by $(1/\Delta_{\rm lin} s)$ in advance, to eliminate the division in Eq.~\eqref{eq:idx_map_lin}, thus improving the overall efficiency of the pair counting algorithm.
Eventually, we need only three operations for the evaluation of bin index for each valid pair, which are square root, subtraction, and floor.

Similarly, the index of squared distance $s^2$ for logarithmic bins in the range of $[s_{\rm min}, s_{\rm max})$ can be obtained by
\begin{equation}
\mathfrak{i}_{\rm log} (s^2) = \left\lfloor \frac{ \frac{1}{2} \log s^2 - \log s_{\rm min} }{ \Delta_{\rm log} s } \right\rfloor ,
\quad s_{\rm min}^2 \le s^2 < s_{\rm max}^2 .
\label{eq:idx_map_log}
\end{equation}
Here, $\Delta_{\rm log} s$ is the width of the bins on logarithmic scale.
Again, we can rescale all coordinates and histogram bins to further improve the efficiency. For instance, with a rescaling factor of $(1 / s_{\rm min})$, the $(\log s_{\rm min})$ term in Eq.~\eqref{eq:idx_map_log} can be omitted. Then, if pre-computing the factor $(2 \Delta_{\rm log} s)^{-1}$, we end up with one logarithm, one multiplication, and one floor for the index mapping.

Though the complexity of index mapping algorithms is only $\mathcal{O} (1)$, which outperforms those of comparison-based methods, the actual computing time depends largely on the efficiency of index calculations.
Actually, a considerable amount of comparisons can be accomplished during the evaluation of logarithm in Eq.~\eqref{eq:idx_map_log}.
Therefore, histogram update algorithms based on index mapping functions are not necessarily faster than methods described in Sect.~\ref{sec:hist_comp}, especially when $N_{\rm bin}$ is small.
To make the constant-time complexity effective, we need more efficient index mapping methods than the direct function evaluations, not to mention the limited numerical precision of these functions.

\subsubsection{Index lookup tables}
\label{sec:hist_lookup}

A common way of accelerating the evaluation of a numerical function is to lookup pre-computed values from a table.
This technique can be very efficient if the domain of the function is discrete and reasonably small.
In general, index mapping functions for histogram updates do not fulfil this condition, as the squared distances can be of any value inside $[s_{\rm min}^2, s_{\rm max}^2)$.
Nevertheless, when the edges of histogram bins are integers, it is only the integer part of a squared distance which determines the index of the histogram bin.
In this case, we can create an index lookup table with the keys being the integer parts of all possible squared distance values.
The index mapping process can then be completed by truncating squared distances and looking up indices in the table.
Moreover, the efficiency of this method can benefit from the data locality with the tree structures discussed previously, which reduces the cache-miss rate of table lookup. 

This method is also applicable if all the histogram bin edges can be converted into integers by a common rescaling factor, as it is permissible to rescale histogram bins together with coordinates of data points.
This is actually a common scenario in practice.
For instance, given equally spaced separation bins with $s_{\rm min} = 0$, the rescaling factor that converts all bin edges into integers is simply the inverse of the bin width.
However, since the length of the lookup table is
\begin{equation}
N_{\rm table} = \lfloor s_{\rm max}^2 \rfloor - \lfloor s_{\rm min}^2 \rfloor \, ,
\label{eq:hist_ntable}
\end{equation}
when the (rescaled) distance range is wide, the table may be too large to fit in the CPU caches. As the result, the lookup efficiency can be downgraded significantly due to the expensive memory accesses.
One solution to this problem is to rescale histogram bins by a factor that is smaller than 1. But then the bin edges are not guaranteed to be integers.
Considering also cases that the bin edges cannot all be converted into machine-representable integers, an index lookup algorithm that does not rely on integer bin edges is necessary.

For non-integer bin edges, we have to take care of non-injective lookup table entries. This is because squared distances belonging to different separation bins may share the same integer part.
In this case, we can record the index ranges for non-injective entries, and use a comparison-based method to further identify the exact index for a given squared distance.
In practice, we use the reverse traversal algorithm (see Sect.~\ref{sec:hist_comp}) due to its simplicity.
It is worth noting that this hybrid index lookup method is able to deal with separation bins with arbitrary bin edges and widths, as long as the bins are continuous.

The efficiency of this method depends on the rescaling factor of histogram bins.
When the factor is small, there is a higher chance of encountering non-injective table entries, which requires further comparisons that are relatively slow.
In contrast, big rescaling factors yield large tables that may increase the cache-miss rate.
In principle, the optimal rescaling factor depends on the CPU cache sizes, and should be estimated through benchmarks.

\subsubsection{Comparisons of the histogram update algorithms}
\label{sec:hist_benchmark}

In order to compare the performances of different histogram update algorithms discussed so far, and choose the optimal table size for the hybrid index lookup method, we perform a series of benchmark tests on \textsf{Haswell} CPUs with squared distance values sampled randomly following Appendix~\ref{sec:dist_s2}.
In particular, the squared distances are sampled in the range of $[0,200^2)\,h^{-2}\,{\rm Mpc}^2$.
We examine both linear and logarithmic separation bins, which are the most commonly used binning schemes in practice, with the $s$ ranges of $[0,200)$ and $[0.1,200)\,h^{-1}\,{\rm Mpc}$ respectively.
To inspect the $N_{\rm bin}$ dependences of the algorithms, we further test two different numbers of histogram bins, 20 and 200, for both binning schemes.
Note that some of the algorithms require rescaling of squared distances, which can be achieved by pre-processing the coordinates of all data points in reality. The computational cost of this pre-processing step is $\mathcal{O} (N)$, which is generally much smaller than that of the histogram update process with $\mathcal{O} (N^2)$ pairs. Thus, the costs of histogram update routines we report do not include those for rescaling separations.

\begin{figure}
\centering
\includegraphics{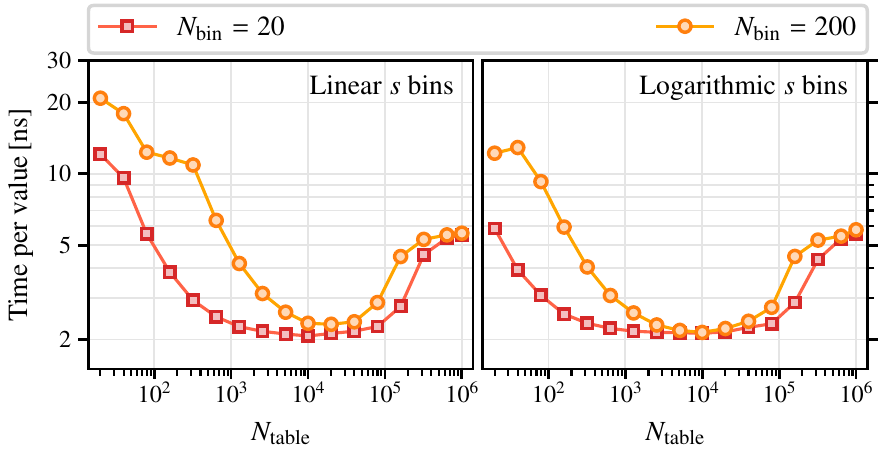}
\caption{Performances of the separation histogram update routine based on the hybrid index lookup algorithm with different lookup table sizes.
The execution time is measured with $6.4 \times 10^9$ randomly sampled squared distances in the range of $[0,200^2)\,h^{-2}\,{\rm Mpc}^2$.
Both linear and logarithmic separation bins are tested, with the $s$ ranges of $[0,200)$ and $[0.1,200)\,h^{-1}\,{\rm Mpc}$ respectively.
There are also two different numbers of bins, 20 and 200, for both binning schemes.}
\label{fig:hist_ntable}
\end{figure}

Given $6.4 \times 10^9$ random squared distances, the execution times of the hybrid index lookup algorithm with different lookup table sizes and histogram bins are shown in Fig.~\ref{fig:hist_ntable}.
It can be seen that a table with $\sim 10^4$ entries is always near-optimal, regardless of the separation bin configurations.
Considering the fact that the indices of histogram bins can be represent by 8- or 16-bit integers in most cases, the memory cost of a table with $\sim 10^4$ entries is around 10 to 20\,KB, which fits in the level-1 (L1) cache of most modern CPUs for supercomputers. This explains the optimality of the table size.
Actually, the optimal histogram update cost per squared distance value is around 2\,ns for all the separation bin configurations in Fig.~\ref{fig:hist_ntable}, which corresponding to barely $\sim 5$ \textsf{Haswell} CPU cycles, so slightly larger than the 4-cycle latency of L1 cache accesses \citep[][]{FogManual}.
This means that we achieve almost the maximum theoretical efficiency for histogram updates.
Thus, we choose always separation rescaling factors that yield $N_{\rm table} \sim 10^4$ for the hybrid index lookup algorithm hereafter.

\begin{figure}
\centering
\includegraphics{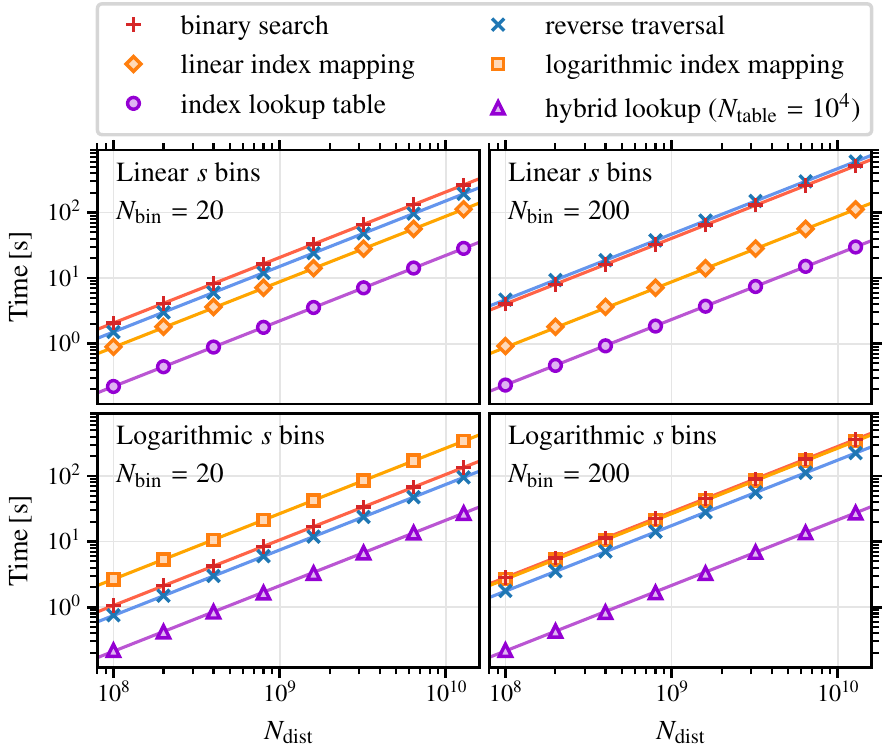}
\caption{Execution times of various histogram update algorithms for different histogram bins and numbers of squared distances sampled in the range of $[0,200^2)\,h^{-2}\,{\rm Mpc}^2$.
The separation ranges of the linear and logarithmic bins are $[0,200)$ and $[0.1,200)\,h^{-1}\,{\rm Mpc}$ respectively.
Two different numbers of bins are also tested.
Solid lines show best-fitting straight lines with constant execution time per squared distance.}
\label{fig:hist_compare}
\end{figure}

We further compare the performances of different histogram update algorithms, with the same input squared separation sequences and histogram bins. The results are presented in Fig.~\ref{fig:hist_compare}.
We do not use the hybrid method for linear separations bins, as the bin edges are integers and lookup tables are directly applicable.
In all cases, the execution time scales linearly with $N_{\rm dist}$, the number of squared distances sampled.
It is not surprising that the comparison-based methods -- binary search and reverse traversal -- are sensitive to both the binning scheme and number of separation bins.
The performances of index mapping algorithms also depend largely on the binning schemes, but not on $N_{\rm bin}$. This can be explained by the different costs of the index mapping functions.
It turns out that the linear index mapping method expressed by Eq.~\eqref{eq:idx_map_lin} is faster than the comparison-based methods for the examined $N_{\rm bin}$ values; while the logarithmic mapping shown in Eq.~\eqref{eq:idx_map_log} is generally less efficient, especially when compared to the reverse traversal algorithm.
In contrast, the index lookup algorithms are insensitive to the configurations of separation bins, and outperform all the other methods in all the cases presented here.
In fact, the lookup cost for each squared distance value is always $\sim 2\,$ns on average.

We then implement the index lookup methods in \fcfc{} for pair counting, due to their high efficiency and the ability of dealing with arbitrary separation bins.
In particular, for linear separation bins, we compute the smallest positive factor that converts both edges of the first bin into integers. Given the separation ranges rescaled by this factor, if the $N_{\rm table}$ expressed by Eq.~\eqref{eq:hist_ntable} is $\lesssim 3 \times 10^4$, we use the index lookup table for integer bin edges directly. For all the other cases -- either the $N_{\rm table}$ computed in this way is too large, or the separation bins are not evenly spaced -- we rely on the hybrid index lookup method with $N_{\rm table} \sim 10^4$.
In order to eliminate potential numerical errors due to the rescaling, we always choose a rescaling factor that is a power of the radix used by floating point representations, and yields a lookup table size that is closest to $10^4$.
In this case, the rescaling changes only the exponent of almost all floating-point numbers, so the mantissas are untouched and no additional numerical errors are introduced.
Once the factor is chosen, we rescale all histogram bins and coordinates of data points accordingly.


\section{Parallelisation}
\label{sec:parallel}

Modern multi-core vector processors are able to run multiple independent instructions simultaneously on different pieces of data.
HPC clusters are usually equipped with hundreds or thousands of such CPUs.
To make full use of the computing facilities, one needs to break down the computational task into similar sub-tasks, and make use of different levels of parallelisms.

\subsection{SIMD}
\label{sec:simd}

Single instruction, multiple data (SIMD) refers to a type of data-level parallelism, which permits operations of multiple data (i.e., a `vector') with a single instruction.
For instance, most of the modern x86 CPUs support Advanced Vector Extensions (AVX), which provides 256-bit registers for 8 single-precision or 4 double-precision floating point numbers to be processed simultaneously.
There are also a number of CPUs that supports Advanced Vector Extensions 2 (AVX2) -- an extension of AVX with the same register width but more instructions -- or even AVX-512, which permits 512-bit SIMD operations.
Note that AVX-512 consists of multiple extension sets.
We focus on AVX-512 Foundation (AVX-512F) in this work, as it is available for all AVX-512 implementations and sufficient for our application.

SIMD is potentially able to boost the performance of a pair counting code, as distances between different pairs of data points can be evaluated at once, which has to be processed for each individual pair with the conventional sequential (also known as `scalar') approach.
Thus, the traversal of points on pairs of tree nodes can be largely accelerated.
In contrast, SIMD does not help much the data pruning process, as the maintenance of the dual-node stack (see Sect.~\ref{sec:dual_tree}) cannot be parallelised with vector operations.
In this case, larger tree nodes and fewer node comparisons are preferred for better overall pair counting efficiency, so the optimal leaf node capacity may change with different register widths.

\begin{figure}
\centering
\includegraphics{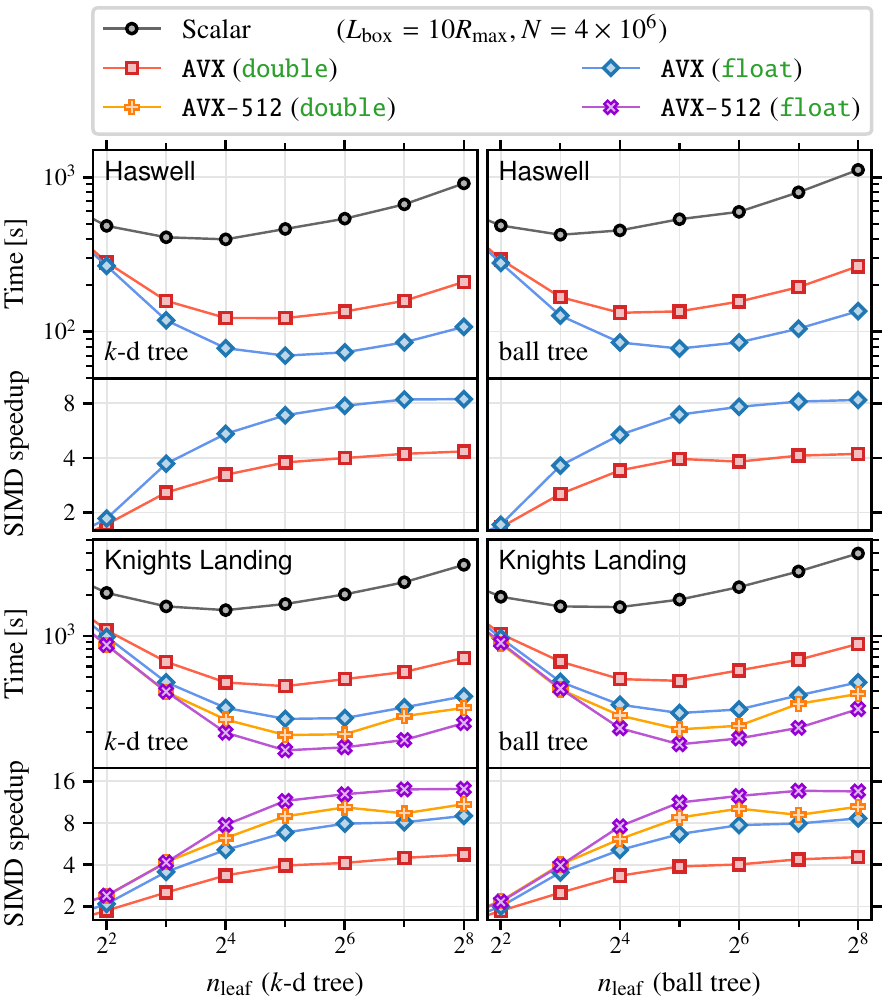}
\caption{Execution time of the scalar, AVX-vectorised, and AVX-512-vectorised pair counting routines based on both $k$-d and ball trees, with different capacities of leaf nodes, for a periodic uniform random sample. Results on both \textsf{Haswell} and \textsf{Knights Landing} CPUs are shown. The speedup is measured as the ratio of the computing time of the scalar code, to that of the vectorised counterpart.}
\label{fig:struct_simd}
\end{figure}

To explore the optimal $n_{\rm leaf}$ for $k$-d tree and ball tree with AVX and AVX-512, we perform a new set of benchmarks with both the scalar and vectorised dual-tree algorithms, on both the \textsf{Haswell} and \textsf{Knights Landing} CPUs (see Appendix~\ref{sec:benchmark_setting}).
In particular, we check both single- and double-precision arithmetics, by using the \texttt{float} and \texttt{double} data types in C progamming language, as illustrated in Fig.~\ref{fig:struct_simd}.
Similar to the tests in Sect.~\ref{sec:data_struct}, we measure the execution time of the pair counting algorithm which reports the number of pairs with separations below $R_{\rm max}$, for $4 \times 10^6$ uniformly distributed random points in a cubic box with the side length of $L_{\rm box} = 10 R_{\rm max}$.
Since it is shown previously that the optimal $n_{\rm leaf}$ is not sensitive to the specifications of the input samples, we do not vary the box size, nor the number of data points here.
Fig.~\ref{fig:struct_simd} shows that with SIMD, $n_{\rm leaf} = 32$ is near optimal for almost all cases.
Moreover, when $n_{\rm leaf} \gtrsim 64$, the theoretical maximum speedups with SIMD are achieved.
For instance, AVX is able to process 4 double-precision numbers at once, and the actual speedups of the AVX-vectorised algorithms are indeed $\sim 4$ with respect to the scalar counterparts.
In fact, the speedup can be larger than the number of floating-point numbers processed simultaneously. This is possibly due to additional efficiency boosts with the fused multiply–add (FMA) instructions that are available with most modern SIMD implementations.

Our histogram update process, however, may or may not benefit from SIMD.
On one hand, the index lookup methods are sufficiently fast that the access of CPU caches may have become the bottleneck (see discussions in Sect.~\ref{sec:hist_benchmark}).
On the other hand, AVX does not provide instructions for reading lookup tables and maintaining histograms.
In this case only the floor operation can be vectorised; while the rest of the histogram update process has to be implemented in scalar.
The more recent AVX2 instruction provides the \texttt{gather} operation, which loads multiple elements from non-contiguous memory locations, and can be potentially useful for loading lookup tables and histogram counts.
Nevertheless, there is still no instruction for the update of histogram with AVX2.
It is only with AVX-512 that both \texttt{gather} and \texttt{scatter} operations are available, where \texttt{scatter} stores multiple data at different memory locations at once.
Therefore, AVX-512 permits a full vectorisation of our histogram update algorithm.

\begin{figure}
\centering
\includegraphics{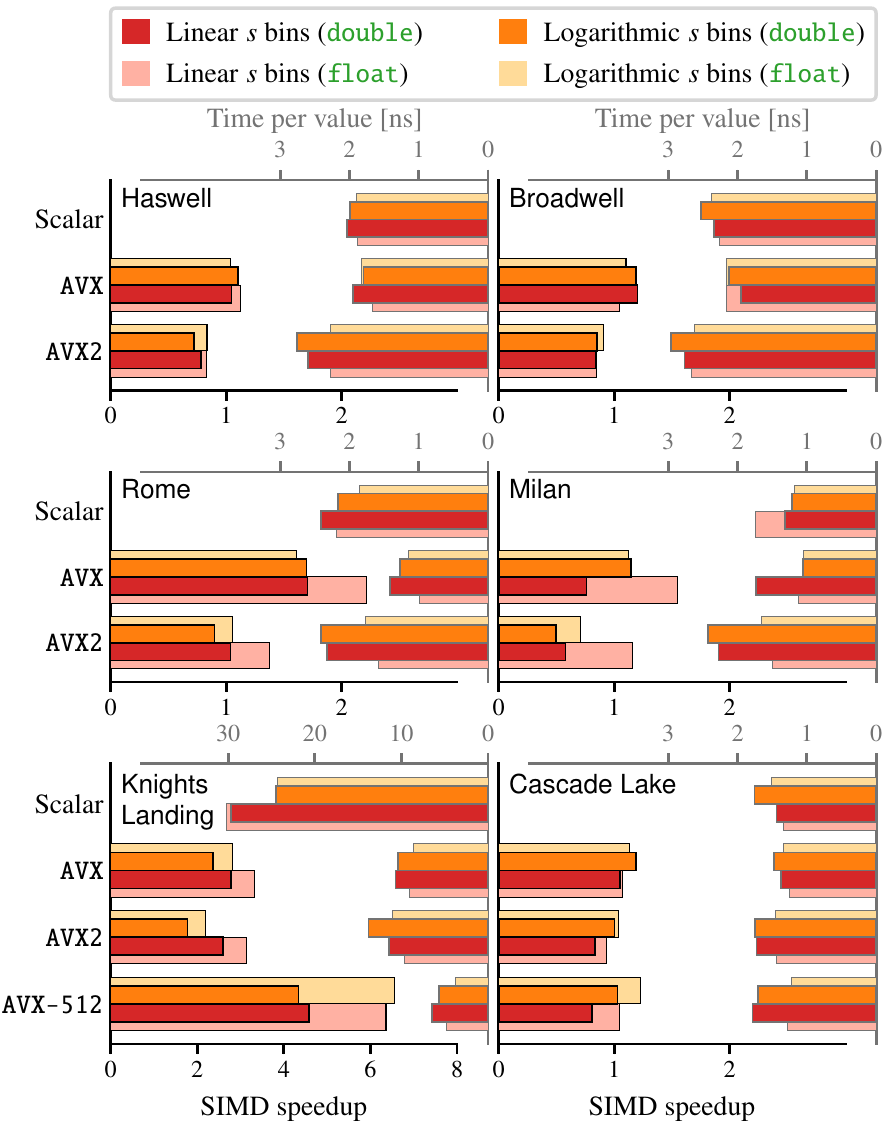}
\caption{Execution time of the histogram update algorithms measured upon $10^{10}$ random squared distances, as well as speedups of the vectorised algorithms with respect to the scalar counterparts on different CPUs, with different histogram bin settings and precisions of floating point numbers.}
\label{fig:hist_simd}
\end{figure}

We then vectorise the histogram update process with different SIMD instruction sets, and perform benchmarks on a number of different CPUs using $10^{10}$ randomly generated squared separation sequences, with the same binning schemes in Sect.~\ref{sec:hist_benchmark}. 
Note that for AVX-512 we maintain private histograms for individual vector elements to avoid conflicts, rather than relying on the Conflict Detection Instructions \citep[AVX-512CD; which are used by][]{CORRFUNC}.
In this way we eliminate costs due to conflict detection and branching by trading off memory usage.
The averaged processing time of each squared separation value, as well as the speedups of the vectorised versions with respect to the scalar counterparts are illustrated in Fig.~\ref{fig:hist_simd}.
The improvements with SIMD are almost always marginal, except for the \textsf{Knights Landing} CPU.
This can be explained by the limits of cache throughputs.
After all, for most of the CPU tested, the cost of processing one square distance value is barely few nanoseconds with the scalar code.
Moreover, with AVX the main components of the histogram update algorithm are not vectorised.
The inclusion of \texttt{gather} instruction alone with AVX2 turns out to be harmful to the efficiency of index lookups, possibly because the algorithm is not fully vectorised, and there are additional micro-operations than memory loads \citep[see Chapter~15,][]{IntelManual}.
The index lookup algorithm can be accelerated significantly by AVX-512 on the \textsf{Knights Landing} CPU, while for \textsf{Cascade Lake} the performances of the vectorised and scalar codes are very similar.
It shows that AVX-512 is only useful when the histogram update procedure is significantly slower than the latency of cache access.
Given these benchmark results, \fcfc{} makes use of \texttt{gather} only when \texttt{scatter}, or AVX-512, is available.
This does not mean that AVX2 is useless, as we benefit from the handy vectorised integer arithmetics introduced by AVX2.

\begin{figure}
\centering
\includegraphics{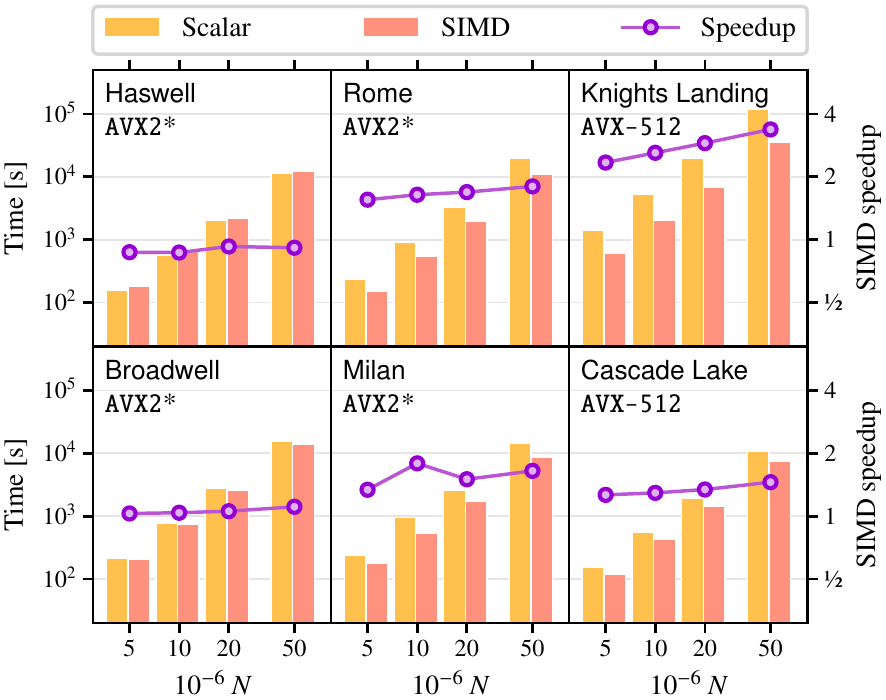}
\caption{Bars show the execution time of the scalar and vectorised versions of \fcfc{} on different CPUs, for the full pair counting procedure with 200 linear $s$ bins in $[0,200)\,h^{-1}\,{\rm Mpc}$ and 120 $\mu$ bins in $[0,1)$, run upon periodic random samples in a cubic box with the side length of $3\,h^{-1}\,{\rm Mpc}$.
Purple lines indicate the speedups of the SIMD-parallelised versions with respect to the scalar counterparts.
`\texttt{AVX2}*' indicates AVX2 but excluding the \texttt{gather} instructions.}
\label{fig:para_simd}
\end{figure}

To further examine whether or how much SIMD is beneficial to the full pair counting procedure, including both the distance evaluations and histogram update, we compare the entire runtime of the scalar and vectorised \fcfc{} on different CPUs for auto pair counts upon periodic cubic random catalogues with the box size of $3\,h^{-1}\,{\rm Gpc}$, with 200 linear $s$ bins in $[0,200)\,h^{-1}\,{\rm Mpc}$ and 120 $\mu$ bins in $[0,1)$, which is a common setting in practice.
The results are presented in Fig.~\ref{fig:para_simd}.
We conclude that SIMD is generally useful, though the overall improvement can be marginal on certain CPUs.
Thus, we always enable SIMD parallelisation throughout this work.

\subsection{OpenMP}
\label{sec:openmp}

Open Multi-Processing\footnote{\url{https://www.openmp.org}} (OpenMP) is a high-level application programming interface (API) that provides a set of compiler directives, library routines, and environment variables for multi-thread parallelisms with shared memory.
It is usually possible to parallelise a program with high scalability using OpenMP, with little modification of the serial code.
Therefore, multi-threading with OpenMP is generally easy to implement, for taking advantage of multi-core processors. It is thus used extensively in cosmological applications, including pair counting programs \citep[e.g.][]{CUTE, GUNDAM, CORRFUNC}.

However, it is not trivial to parallelise our dual-tree algorithm (see Algorithm~\ref{alg:dual_tree}) with high scalability.
The update of dual-node stack has to be executed by one thread at a time to prevent race conditions. This may result in additional overheads.
It is possible to reform the algorithm as a recursive function, but then there are extra costs due to recurrent function calls and creations of threads for subtasks.
One way to eliminate these expenses is maintaining a private stack on each thread.
To this end, the dual-node stack has to be initialised with multiple elements that can be assigned to different threads and run independently.
In this way, the initialisation and allocation of node pairs are crucial for the load balancing of the parallelised dual-tree algorithm.

In principle, one can run Algorithm~\ref{alg:dual_tree} with a single thread until the dual-node stack is sufficiently large, and then distribute the node pairs to different threads.
However, with the depth-first tree traversal order, node pairs on the stack differ significantly in sizes as the number of points on each node depends mainly on the level (or depth) of the node.
In this case, the word loads of different threads are normally highly unbalanced, which is harmful to the efficiency of the parallelised program.
To circumvent this problem, we rely on the breadth-first tree traversal order for the initialisation of node pairs, which are then stored in a queue rather than a stack.
Thus, after each iteration, node pairs in the queue are all at the same level and consist of similar numbers of data points.
Once the queue is large enough, we distribute the node pairs to different OpenMP threads.
Note however that the work loads are still not perfectly balanced in general, as the numbers of pairs within the query range can vary among different node pairs.
It should be possible to further increase the performance of the parallelised dual-tree algorithm by using better scheduling strategies, such as the work stealing technique \citep[e.g.][]{Blumofe1999}. For instance, the scaling efficiency of the 2PCF algorithm developed by \citet[][]{Chhugani2012} is remarkable even with over 25000 threads\footnote{However, the algorithm of \citet[][]{Chhugani2012} is mainly useful for isotropic 2PCFs with a small number of separation bins, so not general enough for actual cosmological applications.}. We leave relevant investigations to a future work.

\begin{figure}
\centering
\includegraphics{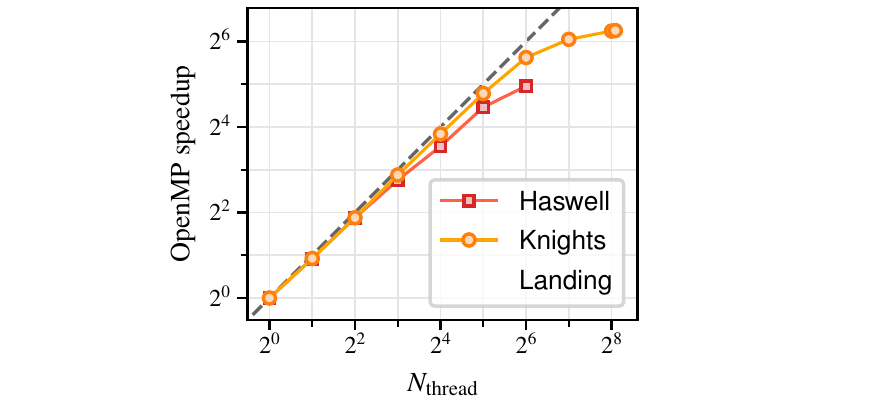}
\caption{Speedups of the OpenMP-parallelised \fcfc{} with respect to the serial version, on \textsf{Haswell} and \textsf{Knights Landing} CPUs with different numbers of OpenMP threads, measured using a periodic cubic random catalogue with $N=5\times10^7$, $L_{\rm box} = 3\,h^{-1}\,{\rm Gpc}$, and with 200 linear $s$ bins in $[0,200)\,h^{-1}\,{\rm Mpc}$ and 120 $\mu$ bins in $[0,1)$.
The dashed line denotes the theoretical maximum speedup.
SIMD is enabled in all cases.}
\label{fig:para_omp}
\end{figure}

The performances of the OpenMP-parallelised \fcfc{} on different CPUs are shown in Fig.~\ref{fig:para_omp}.
The benchmarks are performed with a periodic cubic random sample with $5 \times 10^7$ points, and a box size of $3\,h^{-1}\,{\rm Gpc}$.
Similar to the case in Sect.~\ref{sec:simd}, we measure auto pair counts with 200 linear $s$ bins in $[0,200)\,h^{-1}\,{\rm Mpc}$ and 120 $\mu$ bins in $[0,1)$.
We find that the speedup scales quite well with the number of threads when there are $\lesssim 32$ OpenMP threads.
With more threads the speedups deviates from the theoretical maximum values significantly on both CPUs, possibly because of the non-negligible overheads of maintaining a large number of threads, as well as the imperfect work balancing.
Anyway, the scalability of the OpenMP-parallelised \fcfc{} is reasonably good. Therefore, it is always recommended to enable OpenMP for pair counting tasks with \fcfc{}.

\subsection{MPI}

Message Passing Interface (MPI) is a standard that defines a communication protocol for high-performance parallel computing on distributed memory systems.
It permits multi-process programs that are able to make use of almost all computing resources of a cluster in principle.
In practice, MPI is usually used along with OpenMP.
In this hybrid paradigm, MPI is typically used across computing nodes or sockets of a cluster, while OpenMP is used within nodes or sockets to reduce the communication overhead and memory usage.
Thus, better scalability may be achieved than pure MPI or OpenMP manners.

In fact, our parallelised dual-tree algorithm discussed in Sect.~\ref{sec:openmp} is applicable to MPI parallelism naturally.
After creating the queue with node pairs at the same tree level, one can assign bulks of tasks to different MPI processes, and then repeat the breadth-first tree traversal procedure on each process to generate subtasks for threads if OpenMP is enabled in the meantime.
In this way the pair counting routine is executed independently by different processes and no communication is needed.
Thus, the only additional steps for MPI are the synchronisations of trees and lookup tables among all processes, as well as the gathering of pair counting results at the end.

\begin{figure}
\centering
\includegraphics{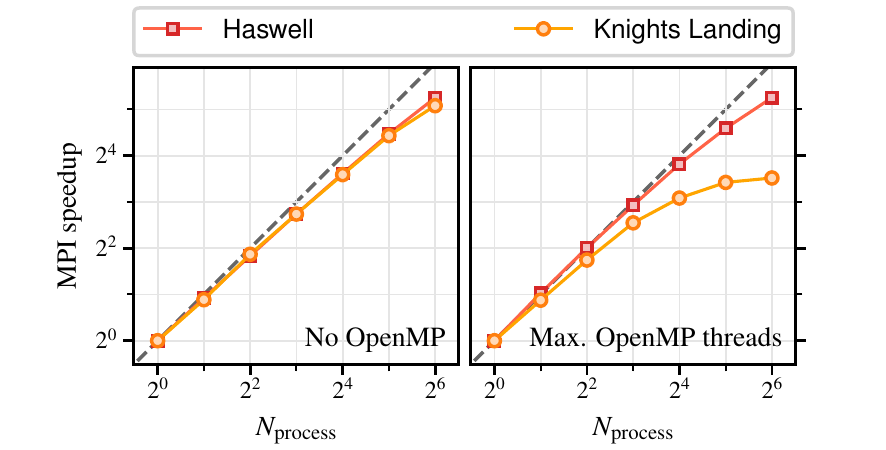}
\caption{Speedups of the MPI-parallelised \fcfc{} with respect to the version without MPI, on \textsf{Haswell} and \textsf{Knights Landing} CPUs with different numbers of MPI processes, measured using periodic cubic random catalogues with $N=5\times10^7$ (\textit{left}) and $5\times10^8$ (\textit{right}), $L_{\rm box} = 3\,h^{-1}\,{\rm Gpc}$, and with 200 linear $s$ bins in $[0,200)\,h^{-1}\,{\rm Mpc}$ and 120 $\mu$ bins in $[0,1)$.
The dashed lines indicate the theoretical maximum speedup.
The \textit{left} panel shows results without OpenMP; while the \textit{right} panel presents results with the maximum available numbers of OpenMP threads. SIMD is enabled in all cases.}
\label{fig:para_mpi}
\end{figure}

The performances of the MPI-parallelised \fcfc{} with and without OpenMP are presented in Fig.~\ref{fig:para_mpi}, for auto pair counts upon periodic cubic random samples with $5\times 10^8$ and $5 \times 10^7$ points respectively, and the same box size and binning scheme as in Sect.~\ref{sec:openmp}.
It can be seen that for \fcfc{} with MPI but without OpenMP, the speedups scale pretty well with the number of MPI processes on both \textsf{Haswell} and \textsf{Knights Landing} nodes.
Actually, the trends are similar to those shown in Fig.~\ref{fig:para_omp}, for which only OpenMP is enabled.
This is expected as we distribute the work loads in the same way.
When enabling OpenMP along with MPI, and running \fcfc{} with the maximum available number of OpenMP threads (64 on \textsf{Haswell} and 272 on \textsf{Knights Landing}), the speedups are basically unchanged on \textsf{Haswell}, but there is a significant degradation of efficiency when the number of processes is $\gtrsim 8$.
This may be due to the fact that small unbalances of work loads become critical with thousands of independent threads running simultaneously. As discussed in Sect.~\ref{sec:openmp}, we leave the exploration into a better work load scheduler to a forthcoming paper.

%

\section{Comparison with related work}
\label{sec:comp_corrfunc}

To see whether \fcfc{} is useful in practice, it is important to run it with real-world applications and compare the efficiency against related pair counting tools.
In fact, \citet[][]{CORRFUNC} have performed extensive benchmarks with a number of different pair counting codes, and concluded that \corrfunc{} outperforms all the other publicly available tools they have checked -- including
\texttt{SciPy cKDTree}\footnote{\url{https://docs.scipy.org/doc/scipy/reference/generated/scipy.spatial.cKDTree.html}} \citep[][]{SCIPY},
\texttt{Scikit-learn KDTree}\footnote{\url{https://scikit-learn.org/stable/modules/generated/sklearn.neighbors.KDTree.html}} \citep[][]{SKLEARN},
\texttt{kdcount}\footnote{\url{https://doi.org/10.5281/zenodo.1051242}},
\texttt{Halotools} \citep[][]{HALOTOOLS},
\texttt{TreeCorr} \citep[][]{TREECORR},
\texttt{CUTE} \citep[][]{CUTE},
\texttt{MLPACK RangeSearch} \citep[][]{MLPACK},
and \texttt{SWOT}\footnote{\url{https://github.com/jcoupon/swot}}
-- for auto pair counts with logarithmic bins upon simulation catalogues with $\gtrsim 10^5$ objects in a cubic volume of $1100^3\,h^{-3}\,{\rm Mpc}^3$.
Thus, for simplicity, we compare \fcfc{} (version 1.0.1\footnote{\url{https://github.com/cheng-zhao/FCFC/releases/tag/v1.0.1}}) only with \corrfunc{} (version 2.4.0\footnote{\url{https://github.com/manodeep/Corrfunc/releases/tag/2.4.0}}) in this work.

Since the most expensive tasks in reality are usually random--random pair counts, we focus only on auto pair counts with random catalogues.
Due to the differences in boundary periodicity and line-of-sight for pair counting with simulation and observational data (see Sect.~\ref{sec:discuss_struct}), we examine two sets of randoms:
\begin{enumerate*}[label=(\arabic*), itemjoin={{, }}, itemjoin*={{, and }}, labelwidth=0pt]
\item
$5 \times 10^8$ uniformly distributed random points in a cubic box with side length $3\,h^{-1}\,{\rm Gpc}$ to mimic the random catalogue for a periodic simulation
\item
the actual random samples for the BOSS DR12 data\footnote{We merge `\texttt{random0\_DR12v5\_CMASSLOWZTOT\_North.fits.gz}' and `\texttt{random1\_DR12v5\_CMASSLOWZTOT\_North.fits.gz}' in \url{https://data.sdss.org/sas/dr12/boss/lss/}, to form a random sample with $\sim 9 \times 10^7$ objects.}, with weights enabled for pair counting.
\end{enumerate*}
These two random catalogues are further down-sampled randomly, for benchmarks with smaller datasets.
For all catalogues we perform pair counts with the following binning schemes:
\begin{enumerate}[label=(\arabic*)]
\item
40 linear $s$ bins in $[0,200)\,h^{-1}\,{\rm Mpc}$ and 20 linear $\mu$ bins in $[0,1)$;
\item
\label{enu:bins}
200 linear $s$ bins in $[0,200)\,h^{-1}\,{\rm Mpc}$ and 120 linear $\mu$ bins in $[0,1)$;
\item
40 logarithmic $s$ bins in $[0.1,200)\,h^{-1}\,{\rm Mpc}$ and 20 linear $\mu$ bins in $[0,1)$;
\item
40 linear $\sigma$ bins in $[0,200)\,h^{-1}\,{\rm Mpc}$ and 200 linear $\pi$ bins\footnote{The numbers of $\sigma$ and $\pi$ bins are different, as \corrfunc{} only allows linear $\pi$ bins with a width of $1\,h^{-1}\,{\rm Mpc}$.} in $[0,200)\,h^{-1}\,{\rm Mpc}$.
\end{enumerate}

\begin{figure*}
\centering
\includegraphics{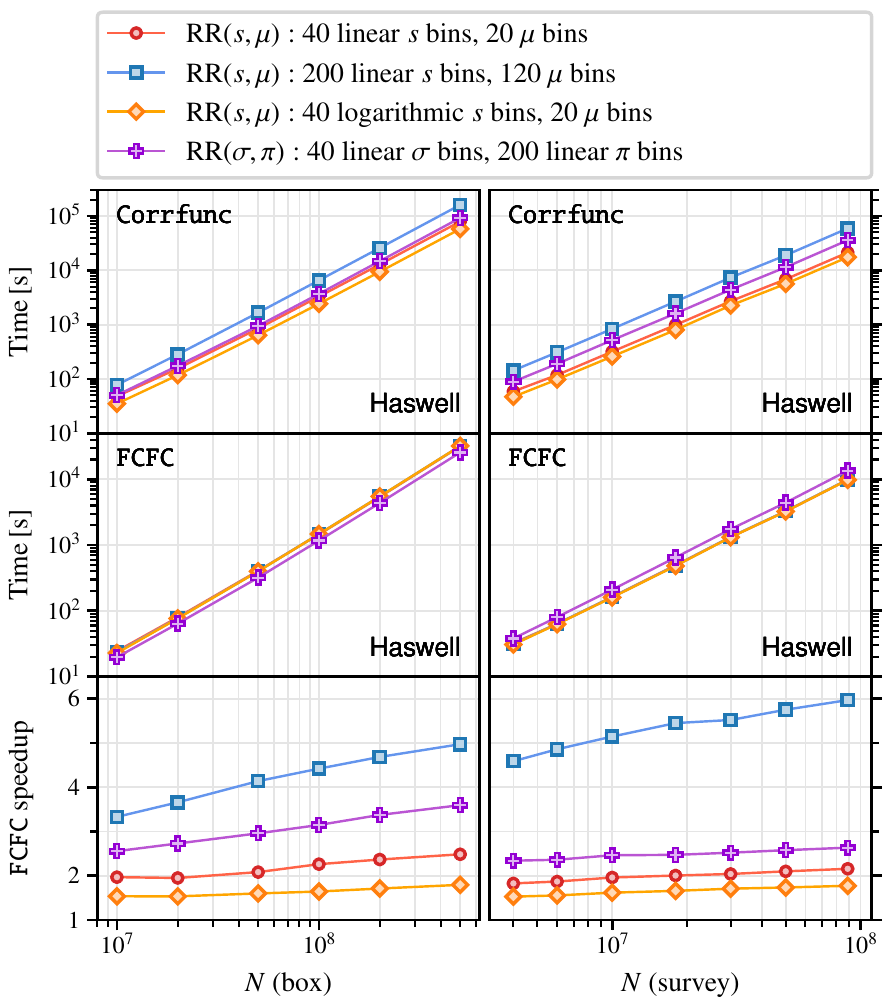}
\hspace{5pt}
\includegraphics{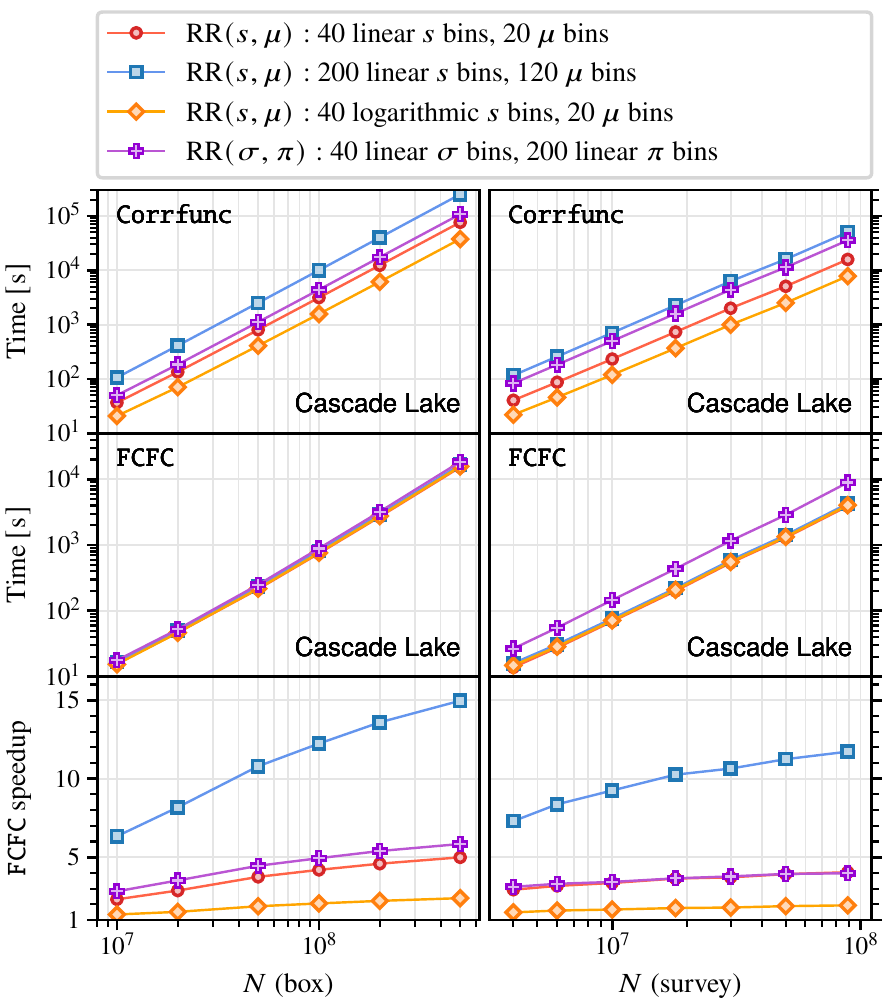}
\caption{Performances of \fcfc{} and \corrfunc{} for auto pair counts with periodic and survey-like random samples with different numbers of points. OpenMP and SIMD parallelisms are both enabled. The codes are run on entire nodes with all cores of \textsf{Haswell} and \textsf{Cascade Lake} CPUs.}
\label{fig:corrfunc}
\end{figure*}

The performances of \fcfc{} and \corrfunc{} on the \textsf{Haswell} and \textsf{Cascade Lake}\footnote{We do not test with \textsf{Knights Landing} CPUs as \corrfunc{} requires more advanced AVX-512 instructions than those available on \textsf{Knights Landing}.} nodes with double-precision arithmetics are shown in Fig.~\ref{fig:corrfunc}.
Here, we enable OpenMP and SIMD for both codes\footnote{\corrfunc{} is not MPI-parallelised.} and run them on an entire computing node with all resources available, that is, 64 threads with AVX2 on \textsf{Haswell}, and 36 threads with AVX-512 on \textsf{Cascade Lake}.
Note that the pair counting results from the two codes are identical, so we only compare their efficiencies here.
One can see that \fcfc{} is faster than \corrfunc{} for all cases.
For logarithmic bins the speedups of \fcfc{} are relatively small; while with linear bins, especially when the bin counts are large, the speedups can be prominent.
In fact, for the binning scheme \ref{enu:bins}, which is commonly used in practice for the ease of re-binning with different bin widths, \fcfc{} can be 5 and 10 times faster than \corrfunc{} with $\gtrsim 10^8$ objects, on \textsf{Haswell} and \textsf{Cascade Lake} CPUs, respectively.
The speedups are generally consistent with those from the index-lookup algorithms for histogram update (see Sect.~\ref{sec:hist_benchmark}).
Thus, we conclude that the high efficiency of \fcfc{} is mainly due to the novel histogram update algorithm.

%

\section{Conclusions}
\label{sec:conclusion}

We have presented \fcfc{}, a high-performance software package for exact pair counting.
It is highly optimised for cosmological applications, but should be useful for the calculations of all kinds of 2-point correlation functions or radial distribution functions with 3D data.
We focus mainly on the efficiency and scalability of the tool in this paper, but \fcfc{} is also portable, flexible, user-friendly, and applicable to a number of different practical problems, such as the calculation of radial distribution functions in statistical mechanics. A brief guide to the toolkit can be found in Appendix~\ref{sec:guide_fcfc}.

We have compared three different data structures for pair counting applications, i.e., regular grids, $k$-d tree, and a novel variant of ball tree. For the tree structures we make use of an improved dual-tree algorithm for pair counting.
We show that the performance of regular grids is sensitive to the choice of grid size. With a sub-optimal grid size, the efficiency of pair counting procedure can be substantially degraded.
In contrast, the tree-based methods are almost always optimal for a fixed capacity of leaf nodes, thus there is no free parameter for the tree constructions.
When the number of data point is sufficiently large, both trees outperform regular grids regardless of the grid size, albeit the improvements may be marginal for cosmological catalogues with $10^7$ -- $10^8$ objects.
Meanwhile, the efficiencies of the two tree structures turn out to be similar.
Therefore, we implement both tree structures in \fcfc{}.

We have further introduced a new histogram update algorithm based on index lookup tables to speedup the increment of separation bins for correlation functions.
For non-integer bin edges, the lookup table is used together with a comparison-based reverse traversal algorithm to locate histogram bins.
Thus, our index lookup method is applicable to arbitrary binning schemes, including multi-dimensional bins for anisotropic measurements.
According to the comprehensive benchmarks with different practical binning schemes, the index lookup method is shown to be considerably faster than the other commonly used histogram update algorithms for all cases.

Then, we parallelise \fcfc{} with three levels of common parallelisms, i.e., SIMD of vector processors, shared-memory OpenMP, and distributed memory MPI, with which it is possible to make full use of all computing resources of a cluster in principle.
It turns out that the key gredient of \fcfc{}, i.e, the index lookup algorithm for separation bin updates, do not get much benefit from SIMD, as the major bottleneck is likely to be the latency of CPU cache accesses.
Nevertheless, the efficiency of \fcfc{} scales well with the numbers of MPI process and OpenMP threads, as long as the total number of threads does not exceed a few thousands.
When the number of threads is too large, the performance boost due to parallelisation may be downgraded.

Finally, we compare OpenMP- and SIMD-parallelised \fcfc{} and \corrfunc{} with the same amount of computing resources, input catalogues, and binning schemes for pair counting. We find that \fcfc{} is faster than \corrfunc{} for all cases tested.
The speedup is the most prominent with a large number of linear separation bins.
In fact, \fcfc{} can be over 10 times faster than \corrfunc{} on modern AVX-512 CPUs, for catalogues containing $\sim 10^8$ objects, and pair counting with 200 linear $s$ bins and 120 $\mu$ bins, which is a common setting for 2PCF calculations in practice.
Thus, \fcfc{} is a very promising tool for modern and future cosmological clustering measurements.

We shall further extend our methods for more cosmological applications in the future, such as angular and high-order clustering statistics, including in particular 3- and 4-point correlation functions. Approximate methods will also be explored to further speedup the measurements with tolerable errors.
Moreover, we are going to implement more advanced load balancing schemes to further increase the scalability of \fcfc{}, and hopefully make use of GPU acceleration.


\begin{acknowledgements}
I thank Charling Tao, Chia-Hsun Chuang, Daniel Eisenstein, and Lehman Garrison for useful discussions on pair counting algorithms.
This work is supported by the Swiss National Science Foundation (SNF) `Cosmology with 3D Maps of the Universe' research grants 200020\_175751 and 200020\_207379.

\fcfc{} also benefits from a number of open-source projects, such as \texttt{Fast Cubic Spline Interpolation}\footnote{\url{https://doi.org/10.5281/zenodo.3611922}} \citep[][]{Hornbeck2020}, \texttt{MedianOfNinthers}\footnote{\url{https://github.com/andralex/MedianOfNinthers}} \citep[][]{Alexandrescu2017}, and \texttt{sort}\footnote{\url{https://github.com/swenson/sort}}.

Benchmarks in this work are run on the Baobab and Yggdrasil HPC clusters at Universit\'{e} de Gen\`{e}ve (UNIGE), as well as the National Energy Research Scientific Computing Center (NERSC)\footnote{\url{https://ror.org/05v3mvq14}}, a U.S. Department of Energy Office of Science User Facility operated under Contract No. DE-AC02-05CH11231.
\end{acknowledgements}

%
%

\bibliographystyle{aa}
\bibliography{FCFC}

\begin{appendix}


\section{Benchmark specifications}
\label{sec:benchmark_setting}

\begin{table}
\setlength{\tabcolsep}{0.9\tabcolsep}
\centering
\caption{Specifications of CPUs on the computing nodes used for the benchmarks.}
\begin{tabular}{lcccc}
\toprule
CPU name & \makecell{\# of\\sockets} & \makecell{\# of\\threads} & AVX2 & AVX-512 \\
\midrule
\textsf{Haswell}\tablefootnote{\url{https://ark.intel.com/content/www/us/en/ark/products/81060/intel-xeon-processor-e52698-v3-40m-cache-2-30-ghz.html}} & 2 & 64 & Yes & No \\
\textsf{Broadwell}\tablefootnote{\url{https://ark.intel.com/content/www/us/en/ark/products/92981/intel-xeon-processor-e52630-v4-25m-cache-2-20-ghz.html}} & 2 & 20 & Yes & No \\
\textsf{Knights Landing}\tablefootnote{\url{https://ark.intel.com/content/www/us/en/ark/products/94035/intel-xeon-phi-processor-7250-16gb-1-40-ghz-68-core.html}} & 1 & 272 & Yes & Yes \\
\textsf{Cascade Lake}\tablefootnote{\url{https://ark.intel.com/content/www/us/en/ark/products/192443/intel-xeon-gold-6240-processor-24-75m-cache-2-60-ghz.html}} & 2 & 36 & Yes & Yes \\
\textsf{Rome}\tablefootnote{\url{https://www.amd.com/en/product/8761}} & 2 & 128 & Yes & No \\
\textsf{Milan}\tablefootnote{\url{https://www.amd.com/en/product/10906}} & 2 & 256 & Yes & No \\
\bottomrule
\end{tabular}
\label{tab:cpu}
\end{table}

We list the node specifications with different CPU architectures used for the benchmarks in this work in Table~\ref{tab:cpu}.
We rely on the \texttt{gcc} compiler\footnote{\url{https://gcc.gnu.org/}} for all our tests, with the compilation flags \texttt{-O3} and \texttt{-march=native} always enabled.
For the \textsf{Haswell} and \textsf{Knights Landing} nodes the compiler version is 7.5.0; while for all the other nodes the version of \texttt{gcc} is 11.2.0.
For tests with MPI we make use of the Open MPI library\footnote{\url{https://www.open-mpi.org/}} version 4.1.2.

The benchmark codes for different data structures and histogram update algorithms are available at \url{https://github.com/cheng-zhao/FCFC/tree/main/benchmark}.
For all benchmarks in this work, each program is run 12 times independently. The execution time is then reported as the averaged cost of 10 runs after excluding the longest and shortest cases.


\section{Complexities of pair counting algorithms based on different data structures}
\label{sec:complexity_struct}

We analyse the complexity of pair counting processes based on different data structures in a simplified case, in which the data points are distributed uniformly in a 3D periodic cubic box with the side length of $L_{\rm box}$, and the distance range of interest is given by $[0, R_{\rm max})$, with $R_{\rm max} \ll L_{\rm box}$.
Note that this is a realistic and interesting scenario in practice, as the most challenging datasets for pair counting are generally from large periodic simulations.

The complexity of a pair counting algorithm consists of two parts:
\begin{enumerate*}[label=(\arabic*), itemjoin={{, }}, itemjoin*={{, and }}, labelwidth=0pt]
\item
$N_{\rm node}$, the number of tree nodes or grid cells that are visited
\item
$N_{\rm pair}$, the number of pairs of data points that are examined.
\end{enumerate*}
Apparently, in the small-node/cell limit, $N_{\rm node}$ dominates the complexity; while $N_{\rm pair}$ is more relevant for large nodes or cells.
We then estimate both $N_{\rm node}$ and $N_{\rm pair}$ for different data structures.

\subsection{Regular grids}
\label{sec:complexity_grid}

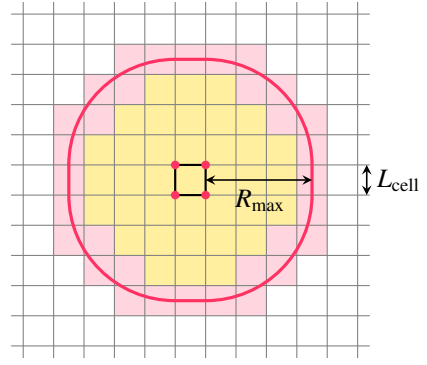
\begin{figure}
\centering
\begin{tikzpicture}
\def \outerpath {(4.5,-2.5) -- (4.5,2.5) -- (3.5,2.5) -- (3.5,3.5) -- (2.5,3.5) -- (2.5,4.5) -- (-2.5,4.5) -- (-2.5,3.5) -- (-3.5,3.5) -- (-3.5,2.5) -- (-4.5,2.5) -- (-4.5,-2.5) -- (-3.5,-2.5) -- (-3.5,-3.5) -- (-2.5,-3.5) -- (-2.5,-4.5) -- (2.5,-4.5) -- (2.5,-3.5) -- (3.5,-3.5) -- (3.5,-2.5) -- cycle}
\def \innerpath {(3.5,-1.5) -- (3.5,1.5) -- (2.5,1.5) -- (2.5,2.5) -- (1.5,2.5) -- (1.5,3.5) -- (-1.5,3.5) -- (-1.5,2.5) -- (-2.5,2.5) -- (-2.5,1.5) -- (-3.5,1.5) -- (-3.5,-1.5) -- (-2.5,-1.5) -- (-2.5,-2.5) -- (-1.5,-2.5) -- (-1.5,-3.5) -- (1.5,-3.5) -- (1.5,-2.5) -- (2.5,-2.5) -- (2.5,-1.5) -- cycle}
\fill[nodebgcolor] \innerpath;
\fill[visitedmem, even odd rule] \outerpath \innerpath ;
\foreach \x in {0,...,11} {
    \draw[split2, gray] (-5.9,\x-5.5) -- (5.9,\x-5.5);
    \draw[split2, gray] (\x-5.5,-5.9) -- (\x-5.5,5.9);
}
\draw[thick] (-0.5,-0.5) rectangle (0.5,0.5);
\filldraw[querypoint] (-0.5,-0.5) circle [radius=2.5pt];
\filldraw[querypoint] (-0.5,0.5) circle [radius=2.5pt];
\filldraw[querypoint] (0.5,-0.5) circle [radius=2.5pt];
\filldraw[querypoint] (0.5,0.5) circle [radius=2.5pt];
\draw[querypoint] (4,0.5) arc[start angle=0, end angle=90, radius=3.5] -- (0.5,4) -- (-0.5,4) arc[start angle=90, end angle=180, radius=3.5] -- (-4,0.5) -- (-4,-0.5) arc[start angle=180, end angle=270, radius=3.5] -- (-0.5,-4) -- (0.5,-4) arc[start angle=270, end angle=360, radius=3.5] -- (4,-0.5) -- cycle;
\draw[stealth-stealth, black, split0] (0.5,0) -- (4,0) node[pos=0.515, below] {$R_{\rm max}$};
\draw[stealth-stealth, black, split0] (5.8,-0.5) -- (5.8,0.5) node[pos=0.5, right] {$L_{\rm cell}$};
\end{tikzpicture}
\caption{Grid cells to be visited (coloured areas) for a reference cell (black square) and an isotropic query range with the radius of $R_{\rm max}$. Yellow regions indicate cells that are entirely inside the query range; while pink zones denote cells intersecting with the boundary of the query range, which is shown in red. The side length of every cell is denoted by $L_{\rm cell}$.}
\label{fig:cell_visit}
\end{figure}

\begin{figure}
\centering
\includegraphics{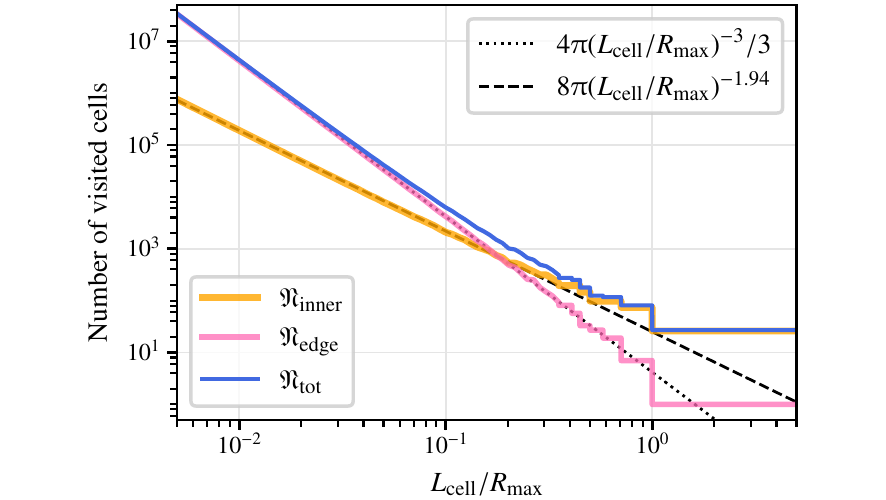}
\caption{The number of regular grid cells to be visited for each  reference cell with side length $L_{\rm cell}$, given a spherical range searching with the maximum distance of $R_{\rm max}$, and a periodic box that is sufficiently large. Here, $\mathfrak{N}_{\rm tot} = \mathfrak{N}_{\rm inner} + \mathfrak{N}_{\rm edge}$, where $\mathfrak{N}_{\rm inner}$ and $\mathfrak{N}_{\rm edge}$ indicate the number of cells that are fully and partially inside the query range, which correspond to the yellow and pink regions in Fig.~\ref{fig:cell_visit}, respectively. The black dashed and dotted lines show analytical formulae that fit well the numerical results in the small-cell limit.}
\label{fig:grid_ncell}
\end{figure}

For cubic datasets, it is obvious that the cells of regular grids are best to be cubes.
In this case, the query range and grid cells to be visited for a single reference cell are illustrated in Fig.~\ref{fig:cell_visit}.
Given the edge length $L_{\rm cell}$ of all grid cells, the number of cells to be visited for any given reference cell, denoted by $\mathfrak{N}_{\rm tot}$, depends solely on $\hat{L}_{\rm cell} \equiv L_{\rm cell} / R_{\rm max}$, as it does not change when rescaling $L_{\rm cell}$ and $R_{\rm max}$ simultaneously with the same factor.
$\mathfrak{N}_{\rm tot}$ can be decomposed into two components:
\begin{equation}
\mathfrak{N}_{\rm tot} = \mathfrak{N}_{\rm inner} + \mathfrak{N}_{\rm edge} ,
\label{eq:ncell_tot}
\end{equation}
where $\mathfrak{N}_{\rm inner}$ and $\mathfrak{N}_{\rm edge}$ indicate the numbers of cells that are fully and partially inside the query range, as shown in yellow and pink in Fig.~\ref{fig:cell_visit}, respectively.
These numbers can be evaluated numerically, and the results are shown in Fig.~\ref{fig:grid_ncell}, together with two empirical analytical formulae that fits well with $\mathfrak{N}_{\rm inner}$ and $\mathfrak{N}_{\rm edge}$ respectively in the small-cell limit.
In particular, when $\hat{L}_{\rm cell} \lesssim 0.5$,
\begin{align}
&\mathfrak{N}_{\rm inner} (\hat{L}_{\rm cell}) \approx
4 \uppi\, \hat{L}_{\rm cell}^{-3} / 3, \label{eq:ncell_inner}  \\
&\mathfrak{N}_{\rm edge} (\hat{L}_{\rm cell}) \approx
8 \uppi\, \hat{L}_{\rm cell}^{-1.94} .
\label{eq:ncell_edge}
\end{align}
In contrast, $\mathfrak{N}_{\rm inner}$ and $\mathfrak{N}_{\rm edge}$ are constants of 1 and 26 respectively when $\hat{L}_{\rm cell} \ge 1$.

Given $N$ data points that are uniformly distributed, with a number density of $\rho = N / L_{\rm box}^3$, the number of pair separations to be computed for the full dataset is a function of $\mathfrak{N}_{\rm tot} (\hat{L}_{\rm cell})$:
\begin{equation}
N_{\rm pair}^{\rm grid} = \rho\, \mathfrak{N}_{\rm tot} (\hat{L}_{\rm cell}) L_{\rm cell}^3  \cdot N 
= N^2 (L_{\rm cell} / L_{\rm box} )^3 \mathfrak{N}_{\rm tot} (\hat{L}_{\rm cell}) .
\label{eq:grid_npair}
\end{equation}
Note in particular that ideally the number of points in cells that are entirely inside the query range can be reported directly. But this is impractical for real-world pair counting problems with multiple separation bins. Therefore we process individual points of these cells anyway (see Sect.~\ref{sec:discuss_struct} for more discussions).
The total number of cells that are visited can be estimated by
\begin{equation}
N_{\rm node}^{\rm grid} = \mathfrak{N}_{\rm tot} (\hat{L}_{\rm cell}) \cdot \min \{ (L_{\rm box} / L_{\rm cell} )^3 , N \} ,
\label{eq:grid_nnode}
\end{equation}
where $(L_{\rm box} / L_{\rm cell} )^3$ is the number of all grid cells, and the term $\min ( L_{\rm box}^3 L_{\rm cell}^{-3}, N )$ indicates an approximation of the number of cells containing data, which reduces to $N$ in the small-cell limit, as most of the cells are empty in this case.

Since $N_{\rm pair}^{\rm grid}$ and $N_{\rm node}^{\rm grid}$ dominate computational costs at the large- and small-cell ends respectively, it is not difficult to find that the complexity of the grid-based pair counting algorithm scales with $\mathcal{O}(L_{\rm cell}^3)$ when $L_{\rm cell} \gtrsim R_{\rm max}$; while it is $\mathcal{O} (L_{\rm cell}^{-3})$ if $L_{\rm cell} \ll R_{\rm max}$.
These relationships are consistent with the measurements shown in Fig.~\ref{fig:grid_benchmark}, where the best-fitting $( a N_{\rm pair}^{\rm grid} + b N_{\rm node}^{\rm grid} )$ curves are also illustrated.
Here, $a$ and $b$ are constants obtained from least-squares fits to the measurements with all different configurations. The agreement between the data and model is good in general, especially for the large- and small-cell ends.

\subsection{\texorpdfstring{$k$}{k}-d tree}
\label{sec:complexity_kdtree}

When constructing the $k$-d tree upon a periodic cubic box with uniform data distribution, the subdivided volumes after space partition are expected to be small cubes the cell size of $(n_{\rm leaf} / \rho)^{1/3}$.
In this case, the number of $k$-d tree leaf nodes with the partitioned volumes intersecting with the query boundaries is close to that of regular grids, which are shown as pink regions in Fig.~\ref{fig:cell_visit}, but with some important differences. 
Firstly, the query range given a reference $k$-d tree node is slightly smaller than that of regular grids, as we measure distances between nodes using their minimum AABBs, which are generally smaller than the corresponding grid cells.
Similarly, it is possible that the AABB of a node does not cross the query range boundary, even if the corresponding subdivided volume intersects with it.
For instance, when there is only a single point on each leaf node, no leaves intersect with the boundary of the query range, as the AABBs reduce to the points, which can only be inside or outside the range.
For both reasons, the number of leaf nodes with their minimum AABBs intersecting with the query boundary is smaller than the prediction of $\mathfrak{N}_{\rm edge} (\hat{n}_{\rm leaf}^{1/3})$,
and may be modelled with an additional term, i.e.
\begin{equation}
\mathfrak{N}_{\rm leaf} = \eta (n_{\rm leaf})\,\mathfrak{N}_{\rm edge} (\hat{n}_{\rm leaf}^{1/3}) ,
\label{eq:ncell_leaf}
\end{equation}
where
\begin{equation}
\hat{n}_{\rm leaf} \equiv n_{\rm leaf}\, \rho^{-1} R_{\rm max}^{-3} .
\end{equation}
When $R_{\rm max}$ is large, the reduction of the query range is not significant. In this scenario, $\eta$ is dominated by the fact that the AABBs of leaf nodes are less likely to intersect with the query range boundaries than regular grids.
The lower limit of $\eta$ is given by the ratio of the AABB volume to that of a grid cell, which is $[ (n_{\rm leaf} - 1 ) ( n_{\rm leaf} + 1 )]^3$ for uniformly distributed points.
For simplicity, we assume
\begin{equation}
\eta (n_{\rm leaf}) \approx \left( \frac{ n_{\rm leaf} - 1 }{n_{\rm leaf}} \right)^3 ,
\label{eq:aabb_ratio}
\end{equation}
which fulfils the condition $\eta(1) = 0$, and approaches 1 when $n_{\rm leaf}$ is sufficiently large.

Since the tree structure is self-similar, the number of node separation evaluations, $N_{\rm node}^{k\textrm{-d}}$, can be solved recursively.
For instance, for a $k$-d tree that contains at most $n_{\rm leaf}$ data points per leaf node, with $n_{\rm leaf} > 1$, further dividing the leaves into two parts is as if constructing a new tree with a leaf capacity of $(n_{\rm leaf} / 2)$.
Moreover, if the separation range between two original leaf nodes intersects with the boundary of the query range, the separations between their both children are checked for pair counting with the new tree.
Consequently, we have
\begin{equation}
N_{\rm node}^{k\textrm{-d}} ( \frac{n_{\rm leaf}}{2}) - N_{\rm node}^{k\textrm{-d}} (n_{\rm leaf})
= \frac{ 4 N }{ n_{\rm leaf}} \cdot \mathfrak{N}_{\rm leaf} (\hat{n}_{\rm leaf}^{1/3}) ,
\label{eq:kdtree_nnode_rec}
\end{equation}
where $(N / n_{\rm leaf})$ is an approximation of the total number of leaf nodes for the original $k$-d tree.
Since the number of visited node is only significant when there are lots of nodes, in which case $\hat{n}_{\rm leaf}$ is small, we consider here only the small-cell end of $\mathfrak{N}_{\rm leaf}$.
Given also Eqs.~\eqref{eq:ncell_edge}, \eqref{eq:ncell_leaf}, and \eqref{eq:aabb_ratio}, the right hand side of this recursive equation is a Laurent polynomial of $n_{\rm leaf}$, which yields the following analytical solution:
\begin{equation}
N_{\rm node}^{k\textrm{-d}} \propto 
\left( 15 - \frac{18}{n_{\rm leaf}} + \frac{8.3}{n_{\rm leaf}^2} - \frac{1.3}{n_{\rm leaf}^3} \right)
\cdot \frac{\uppi \, N^{1.65} R_{\rm max}^{1.94}}{n_{\rm leaf}^{1.65} L_{\rm box}^{1.94}} .
\label{eq:nnode_kdtree}
\end{equation}

When considering the number of pair separations that are evaluated during the dual-tree pair counting process, one can count the number of leaf nodes that are not entirely outside the query range, even though the algorithm may terminate without visiting all leaves.
This is because for each node of the tree, the associated dataset is the union of the ones on all the corresponding descendant leaf nodes.
Therefore, the total number of pair separations computed for the full dataset is
\begin{equation}
\begin{aligned}
N_{\rm pair}^{k\textrm{-d}} &= n_{\rm leaf}\, ( \mathfrak{N}_{\rm inner} + \mathfrak{N}_{\rm leaf} ) \cdot N \\
&= n_{\rm leaf} N \left[ \mathfrak{N}_{\rm inner} (\hat{n}_{\rm leaf}^{1/3}) + (n_{\rm leaf} - 1)^3 n_{\rm leaf}^{-3} \, \mathfrak{N}_{\rm edge} (\hat{n}_{\rm leaf}^{1/3}) \right] .
\end{aligned}
\label{eq:npair_kdtree}
\end{equation}

Therefore, in the large-node limit, the complexity of the pair counting algorithm based on $k$-d tree scales with $\mathcal{O} (n_{\rm leaf}^{0.35})$; while it is a Laurent polynomial of $n_{\rm leaf}$ for small tree nodes (see Eq.~\eqref{eq:nnode_kdtree}).
The best-fitting $( a N_{\rm pair}^{k\textrm{-d}} + b N_{\rm node}^{k\textrm{-d}} )$ curves are shown in Fig.~\ref{fig:kdtree_benchmark}, where the constants $a$ and $b$ are obtained by least-squares fits to all the measurements. The theoretical complexity agrees remarkably well with the data for almost all cases.

Since ball tree is a similar data structure as $k$-d tree, especially for cubic periodic boxes, the derivations for $k$-d tree should work for ball tree as well, albeit the relationship in Eq.~\eqref{eq:aabb_ratio} may be slightly different due to a different representation of the node bounding volume.
We then fit the theoretical complexity from Eqs.~\eqref{eq:nnode_kdtree} and \eqref{eq:npair_kdtree} to the measurements shown in Fig.~\ref{fig:balltree_benchmark}, and the agreements turn out to be excellent.


\section{Random sampling of squared pair separations}
\label{sec:dist_s2}

For periodic boxes, Eq.~\eqref{eq:npair_valid} shows that the total number of pairs with separations below $R_{\rm max}$ scales with $R_{\rm max}^3$.
In this case, the probability distribution function (PDF) of pair separations satisfies
\begin{equation}
P (s) \propto s^2 .
\label{eq:pdf_s}
\end{equation}
The goal is to reproduce this distribution with uniform random sequences in the range $[0,1)$, which are the direct outputs of most random number generation algorithms in practice.
Denoting such a random number as $x$, we have then $P(x) = 1$, and needs to find the relation $s(x)$, such that Eq.~\eqref{eq:pdf_s} holds.

When transforming a variable $x$ to $y$, with $y(x)$ being monotonic, the PDFs of $x$ and $y$ satisfies
\begin{equation}
P_y (y) = P_x (x(y)) \left| \frac{{\rm d} x}{{\rm d} y} \right| .
\label{eq:pdf_transform}
\end{equation}
Given this relation, we find
\begin{equation}
s(x) \propto x^{1/3} .
\end{equation}
In other words, to sample randomly squared pair separations in the range $[0,1)$, one barely needs to compute $x^{2/3}$ for uniform random variables $x$ generated in the same range.
To extend the maximum separation to $R_{\rm max}$, the conversion is simply
\begin{equation}
s^2 = x^{2/3} \cdot R_{\rm max}^2 .
\end{equation}

\section{A quick guide to \fcfc{}}
\label{sec:guide_fcfc}

As of version 1.0.1, \fcfc{} supports the following 2PCFs:
$\xi (s)$,
$\xi (s, \mu)$,
$\xi (\sigma, \pi)$,
$\xi_\ell (s)$,
and $w_{\rm p} (\sigma)$,
where
\begin{align}
&\xi_\ell (s) = (2 \ell + 1) \int_0^1 \xi (s,\mu) \mathcal{L}_\ell (\mu) \, {\rm d} \mu , \\
&w_{\rm p} (\sigma) \approx 2 \int_0^{\pi_{\rm max}} \xi(\sigma, \pi) \, {\rm d} \pi .
\end{align}
Here, $\mathcal{L}_\ell$ denotes the Legendre polynomial with order $\ell$.
The correlation function estimator is user-defined and can be arbitrary.
It accepts both periodic and non-periodic input catalogues in ASCII text, FITS, and HDF5 formats.
In particular, the supports of FITS and HDF5 formats require the \texttt{CFITSIO}\footnote{\url{https://heasarc.gsfc.nasa.gov/fitsio/}} and \texttt{HDF5}\footnote{\url{https://www.hdfgroup.org/solutions/hdf5/}} libraries.
Apart from the optional libraries for file formats, as well as the OpenMP and MPI libraries for the corresponding parallelisms, \fcfc{} does not depend on any other external library. It is fully compliant with the ISO C99\footnote{\url{https://www.iso.org/standard/29237.html}} and IEEE POSIX.1-2008\footnote{\url{https://ieeexplore.ieee.org/document/4694976}} standards. Therefore, \fcfc{} can be easily compiled with most modern C compilers and operating systems.

Specifications of a pair counting task can be passed to \fcfc{} via either a configuration file or command line options. 
We introduce here a few handy settings for different practical scenarios.
For instance, the 2PCF of a periodic simulation catalogue is generally measured using the Peebles--Hauser estimator \citep[][]{Peebles1974}:
\begin{equation}
\xi = {\rm DD} / {\rm RR} - 1 ,
\label{eq:xi_ph}
\end{equation}
where RR can be computed analytically.
In this case, the relevant configurations of \fcfc{} can be
\begin{verbatim}
CATALOG       = sim_data.txt
CATALOG_LABEL = D
PAIR_COUNT    = DD
CF_ESTIMATOR  = DD / @@ - 1
\end{verbatim}
Here, \texttt{CATALOG} denotes the filename of the input catalogue, and \texttt{CATALOG\_LABEL} sets the label of this catalogue.
\texttt{PAIR\_COUNT} defines the sources of catalogues forming pairs, so `\texttt{DD}' indicates auto pair counts of the catalogue `\texttt{D}'. Finally, \texttt{CF\_ESTIMATOR} sets the correlation function estimator, where `\texttt{@@}' denotes the analytical RR pair counts. Apparently, the estimator is basically set in the same form as Eq.~\eqref{eq:xi_ph}.

Similarly, given observational luminous red galaxy (LRG) and emission line galaxy (ELG) samples with the filenames `\texttt{LRG\_data.txt}' and `\texttt{ELG\_data.txt}', together with the corresponding random catalogues `\texttt{LRG\_rand.txt}' and `\texttt{ELG\_rand.txt}', respectively, the auto 2PCFs of LRGs and ELGs as well as the cross 2PCFs between LRGs and ELGs can be computed at once with the following \fcfc{} settings:
\begin{verbatim}
CATALOG       = [LRG_data.txt, LRG_rand.txt,
                 ELG_data.txt, ELG_rand.txt]
CATALOG_LABEL = [L, R, E, S]
PAIR_COUNT    = [LL, LR, RR, EE, ES, SS,
                 LE, LS, RE, RS]
CF_ESTIMATOR  = [(LL - 2 * LR + RR) / RR,
                 (EE - 2 * ES + SS) / SS,
                 (LE - LS - RE + RS) / RS]
\end{verbatim}
It can be seen that the Szapudi--Szalay estimator \citep[][]{Szapudi1997} is used for the cross correlation here:
\begin{equation}
\xi^\times = ( {\rm D}_{\rm L} {\rm D}_{\rm E} - {\rm D}_{\rm L} {\rm R}_{\rm E} - {\rm R}_{\rm L} {\rm D}_{\rm E} + {\rm R}_{\rm L} {\rm R}_{\rm E} ) / {\rm R}_{\rm L} {\rm R}_{\rm E},
\end{equation}
where the subscripts `L' and `E' denotes the catalogues for LRGs and ELGs, respectively.

Thanks to the \texttt{libast} library\footnote{\url{https://github.com/cheng-zhao/libast}} embedded in \fcfc{}, human-readable expressions can be used not only for the correlation function estimators, but also numerical values read from the input catalogues.
For example, to compute auto pair counts of the BOSS DR12 combined sample\footnote{\url{https://data.sdss.org/sas/dr12/boss/lss/galaxy_DR12v5_CMASSLOWZTOT_North.fits.gz}} in the redshift range $0.2 < z < 0.5$, one has to use weights to correct for systematics and reduce variance, with the total weight given by \citep[][]{Reid2016}:
\begin{equation}
w_{\rm tot} = w_{\rm FKP} \, w_{\rm sys} \, ( w_{\rm cp} + w_{\rm noz} - 1 ),
\end{equation}
where $w_{\rm FKP}$, $w_{\rm sys}$, $w_{\rm cp}$, and $w_{\rm noz}$ indicate the \texttt{WEIGHT\_FKP}, \texttt{WEIGHT\_SYSTOT}, \texttt{WEIGHT\_CP}, and \texttt{WEIGHT\_NOZ} columns of the data catalogue, respectively.
In this case, \fcfc{} can be configured with
\begin{verbatim}
POSITION  = [${RA}, ${DEC}, ${Z}]
SELECTION = ${Z} > 0.2 && ${Z} < 0.5
WEIGHT    = ${WEIGHT_FKP} * ${WEIGHT_SYSTOT} *
            (${WEIGHT_CP} + ${WEIGHT_NOZ} - 1)
\end{verbatim}
Here, \texttt{\$\{X\}} indicates the column \texttt{X} of the input FITS catalogue.

For more details on the configurations of \fcfc{}, we encourage the readers to check the documentation of the toolkit\footnote{\url{https://github.com/cheng-zhao/FCFC/blob/main/README.md}}.

\end{appendix}

\clearpage\end{CJK*}
\end{document}